\documentclass[12pt]{book}

\usepackage{latexsym}

\usepackage{amsmath}

\begin{document}

\begin{titlepage}
\begin{center}
\large{\bf{Universit\`a degli Studi di Trento\\
Facolt\`a di Scienze Matematiche, Fisiche e Naturali}}
\vspace{1cm}\\
\Large{ \bf{Tesi di Dottorato di Ricerca in Fisica}}\\
\hrulefill \\
\vspace{4cm}
\huge{Two Dimensional Conformal Symmetry and the Microscopic Interpretation of Black Hole Entropy}
\vspace{3cm}\\
\Large{Alex Giacomini}
\vspace{3cm}\\
\hrulefill \\
\normalsize{Dottorato di Ricerca in Fisica XVI Ciclo}
\end{center}
\end{titlepage}

\frontmatter

\chapter{Acknowledgements}
I want to thank my advisor Prof. Luciano Vanzo, who already at the time of my graduation thesis introduced me to black hole physics and
gave me a lot of suggestions for my research work in the last years.\\
I want also thank Prof. Sergio Zerbini and Dr. Valter Moretti for all the useful discussions during those research years.\\
A warm thank goes also to Nicola Pinamonti with whose cooperation a lot of my research work was done.\\
Last but not least I want to thank Marcello Ortaggio,  Marco Caldarelli and Enrico Degiuli for their encouragement and for having made my working place,
with interesting discussions, stimulating and interesting. 

\tableofcontents
 
\mainmatter

\chapter{Introduction}

In this thesis the research work, done by the author in the three years of his Ph.D. study, will be exposed. 
During this time the author worked on the role of two dimensional conformal field theories in the microscopic
interpretation of black hole entropy.\\
The discovery of the thermodynamic properties of black holes happened with the article of Bardeen Carter and Hawking \cite{hawking&bardeen&carter},
where for the first time the ``4 laws of black hole mechanics'' were found and the formal analogy with the 4 laws of thermodynamics
was noticed. Of course at this stage the analogy with thermodynamics was purely formal, because a classical black hole per 
definition is a region of space-time , where ``nothing can escape to infinity''. Therefore classically speaking a black hole 
is ``black'' , i.e. it has no thermal radiation and therefore it has zero temperature.\\
The things changed dramatically when Hawking in 1975 \cite{hawking} discovered that due to quantum effects a black hole emits a thermal
radiation proportional to it's surface gravity.
This effect is  very small, in fact a black hole of 10 solar masses has a temperature of about $10^{-8}K$.
But so the 4 laws of black hole mechanics are not only an analogy to the laws 
of thermodynamics, but are really the laws of thermodynamics applied to black holes. One can therefore associate also 
an entropy to a black hole, proportional to it's horizon area, given by the Bekenstein-Hawking formula \cite{bekenstein}, \cite{bekenstein2}
$S=\frac{A}{4}$ .\\ 
 Now in thermodynamics the  entropy has always a statistical interpretation in terms of microstates 
given the Boltzmann formula $ S=k_B \log N $ . Therefore arises the question which are the the microstates responsible for the 
black hole entropy. One possibility would be, if we consider a black hole as a collapsed star, to take the degrees of freedom of 
matter as responsible for the entropy. But if we compare the entropy of a non collapsed star of one solar mass , which is about $10^{58} k_B$,
with the entropy of a black hole of the same mass , which is about $10^{77} k_B$, we notice that there are 19 magnitudes missing.
Therefore one can conclude that the relevant role for the black hole entropy is played by the gravitational degrees of freedom.
The problem is now that no complete quantum theory of gravity exists at the moment. There exist of course some specific models
for quantum gravity like superstrings \cite{vafa}, loop quantum gravity \cite{ashtekar}
or Sakharov induced gravity \cite{sakharov}, which can successfully compute the entropy of some classes of black holes.
The fact is that one is  able already at classical level to see the thermodynamic properties of black holes. The 4 laws of black
hole thermodynamics are in fact theorems of differential geometry and therefore completely classic. Also the hawking radiation
is found by means of a semiclassical computation, being the the spacetime geometry considered as a fixed background. On the other
side all the different approaches to quantum theory of gravity give always the Bekestein-Hawking entropy as result.
There seems to be a sort of ``universality principle'' regulating the black hole entropy.\\
 One can therefore argue that, as the laws of thermodynamics of black holes are set already at classical level, also the behavior
of the microstates of a black hole is set already at classical level. The question is then, what classical principle can be 
enough strong to force the behavior of gravitational microstates, whose nature is intrinsically quantum mechanical.
An idea would be to use a classical symmetry principle. In fact such a principle would also be inherited, perhaps with quantum corrections,
in a possible quantum theory of gravity in the sense that it's fields would transform under a representation of this symmetry group.    
A case of symmetry group that is enough powerful to determine the behavior of  the density of states is the two dimensional
conformal group. In the quantum case the generators of 2-D conformal transformations form , with respect to the commutator, a Virasoro
algebra \cite{polyakov}  with a central charge. The asymptotic density of states of the system is then completely determined by the value
of the central charge and by the $L_0$ generator by means of the Cardy formula \cite{cardy}.\\
As described here, the arising of a central extension in the generator algebra is a quantum effect due to the normal ordering of 
the creation and annihilation operators. But a central extension of the conformal algebra may already happen at classical level,
e.g. in the canonical representation of the algebra. This fact has already been noticed by Arnold \cite{arnold}.\\
Therefore if one finds, using e.g. a dimensional reduction or choosing a suitable 2-D submanifold , a two dimensional classical
conformal field theory, that describes the black hole and that admits a \emph{classical} central charge, one would be able
to count the microstates via Cardy formula but using a completely classical Virasoro algebra.
This approach is then  completely independent from specific approach of quantum theory of gravity, because it uses only 
classical properties of black holes,therefore it is a sort of ``quantum gravity without quantum gravity''.
On the other side this approach gives an important hint on the form of a possible quantum theory of gravity.\\
The first approach , that is performed in this thesis, is to compute  the Poisson brackets of canonical
diffeomorphisms generators that preserve certain fall off condition of the metric near the event horizon.
the history of this idea is based on an article of Brown \& Henneaux \cite{brown&henneaux} , where the Poisson algebra of the diffeomorphisms generators         
that preserve the asymptotic structure of $AdS_3$ has a computable central charge. Using this result, Strominger \cite{strominger}
computed the entropy of the $ BTZ$ black hole \cite{BTZ}  via Cardy formula. The problem of this approach is that it is limited to the 
$BTZ$ model being embedded in an $AdS_3$ spacetime. The other problem of this approach is, that using  symmetries at 
infinity it is not able to distinguish a black hole from a star.\\
Therefore it seems more natural to use symmetries that preserve certain falloff conditions near the event horizon.
One can notice that all the relevant geometry of the black hole is in the $r-t$ plane, so one can consider only the deformations 
of the $r-t$ plane. Considering only 2-D submanifold this approach becomes valid for any dimension.
Carlip \cite{carlip}, \cite{carlip2} used at first this idea and later also Ghosh et al. \cite{ghosh}. This approaches are very interesting but seem to have some 
technical difficulties \cite{park}, \cite{soloviev}  especially for the Schwarzschild black hole case . \\
In the first part of this thesis therefore we will try to use this approach for Schwarzschild black holes starting directly with
the Schwarzschild geometry and not from the Kerr like in the articles cited before. We will first carefully discuss the nature of the
boundary terms of the generators, which are necessary to make them differentiable. Those boundary terms are responsible for the possible central charge.
This subject is strictly related to the boundary conditions that one puts on the black hole bifurcation e.g. fixing its 
geometry or surface gravity. We will see that it will be more suitable to fix the surface gravity as boundary condition.
Unfortunately it will turn out that in both cases the central charge is zero. This is in accordance with the results in \cite{koga}, \cite{hotta&sasaki}.
Therefore in this approach it is not possible to compute the black hole entropy via Cardy formula.
On the other hand in this approach, also if the central charge was not zero  it would be difficult to understand which 
should be the underlying conformal field theory which gives the correct entropy.\\
In the second part of the thesis therefore we will make another approach,
based on an article written by A.Giacomini \& N.Pinamonti \cite{giacomini&pinamonti}. Instead of restricting us to the $r-t$ plane in
4-D spacetime we will try to find an effective 2-D theory which describes the black hole.
We are always interested in the Schwarzschild black hole that in the cited papers was the most problematic case, although
it is the geometrically most simple black hole. In order to describe a Schwarzschild black hole we make the Ansatz of
spherical symmetry of the metric and perform so the dimensional reduction of the Einstein-Hilbert action.
We obtain 2-D theory with two fields: the metric of the $r-t$ plane and the dilaton field. Now in two dimensions
the metric can always be written in conformally flat form. Using this fact, all the geometry of the $r-t$ plane 
metric is encoded in one field i.e. the conformal factor. Using then a near horizon approximation it is possible 
to rule out the dilaton field and the equation of motion of the conformal factor becomes a Liouville equation. 
The Liouville theory is a 2-D conformal field theory. It is known \cite{jackiw} that the Liouville theory possesses
a classical central charge in the Poisson algebra of the generators. The origin the central charge in the Liouville theory is in the 
nonscalar transformation property of it's field.\\
Due to a normalization problem in our case we have to introduce a cutoff parameter $l$. The central charge
goes then as $1/l$ and tends so to zero in accordance to the results in the first part of the thesis.
But now the $L_0$ generator goes as $l$ and therefore for every finite $l$ we can use the Cardy formula.
In the Cardy formula the central charge and the $L_0$ generator enter as product $cL_0$ and therefore 
the result does not depend on  $l$. We can so safely take the limit for the cutoff that tends to infinity.
Therefore also if the central charge tends to zero we can compute the Entropy of the black hole obtaining 
the Bekenstein-Hawking result.\\
Using this approach we have not only computed the entropy using only classical symmetry properties but
we have also found that the gravitational degrees of freedom alone, without the help of other fields,
are responsible for the entropy. The advantage of this technique, compared to the canonical diffeomorphism generator approach,  
is that we concretely find the conformal field responsible for the entropy, namely the conformal field of the metric.\\
In the last part of the thesis we are going therefore to study more in detail the dynamics of this field.
We will see, that, returning to the original coupled equations of motion of the conformal field and the dilaton,
using the constraints it is possible to integrate the equations of motion in terms of a free field.
Now again using a near horizon approximation we find that this field is proportional to the Liouville field.\\ 
We argue therefore that for a possible quantum approach to the Schwarzschild black hole it may be 
enough to quantize the free field. We will therefore discuss the coupling of this free field with the scalar curvature
and study the improved stress-energy tensor of this coupled theory. The free field coupled to the gravitational field does not transform as a scalar.
As in the Liouville case therefore the Poisson algebra of the charges acquires a classical central charge. 
We have gauge fixed the metric in our classical theory, but in the quantum case the path integral should be independent from the gauge choice.
This is true if the trace of the stress energy tensor is zero also in the quantum case. A possible trace anomaly is proportional to the total 
central charge.  We can therefore fix the constant of the field-curvature coupling in such a way that the classical central charge of the theory plus 
the quantum contribution plus the contribution of the ghosts, introduced in order to gauge fix the path integral eventually cancel.
In this way we can obtain a  consistent quantum  theory describing the black hole microstates.

\chapter{Black holes and classical\\
 conformal symmetry}

\section{Black holes and thermodynamics}

A black hole is a region of spacetime from which nothing can ``escape to infinity''.
Mathematically a black hole $B$ in a spacetime $(M,g)$ is therefore defined as

\begin{equation}
B= M - J^- (\mathcal{I}^+) \; ,                   \label{blholedef}
\end{equation} 
where the quantity $J^- (\mathcal{I}^+)$ is the chronological past of the future null infinity.
The boundary of $B$ is called event horizon $H$.

\begin{equation}
H=\partial J^- (\mathcal{I}^+)  
\end{equation}
Now for the black holes 4 theorems of differential geometry called the ``4 laws of black hole mechanics'' \cite{hawking&bardeen&carter}.
Let us consider the the surface gravity $\kappa$ defined on the event horizon. The zeroth law states that this quantity is constant on
the horizon

\begin{equation}
\kappa = const \; \;  \mathrm{ on \, H}    \; .                                                          \label{law1}
\end{equation}
It is then well known that a black hole is completely characterized by 3 parameters i.e. it's mass $M$ it's angular momentum $J$
and it's electric charge $e$. The first law relates the infinitesimal variation of the black hole mass with the other parameters

\begin{equation}
\delta M = \frac{\kappa}{8\pi} \delta A + \Omega \delta J +\Phi \delta e \; ,                              \label{law2}
\end{equation}
where $A$ is the area of the spatial section of the event  horizon. The second law states that the area $A$ in every physical 
process always increases or at least remains equal.  

\begin{equation}
\delta A = \geq 0 \; \;  \mathrm{in \,  every \,  phys. \, process}           \; .                         \label{law3} 
\end{equation}                                                                    
This means e.g. that if two black holes collide, then the area of the merging black hole is greater or at least equal
to the sum of the areas of the single black holes.
It is interesting to notice that an extremal black hole i.e. with $\kappa=0$ would be unstable in the sense that a small
perturbation would turn it in naked singularity, violating then the cosmic censorship conjecture. 
The third law therefore states that such a black hole state is not achievable in finite number of physical steps       

\begin{equation}
\kappa =0 \; \;  \mathrm {BH \,  not \,  realizable} \label{law4}
\end{equation}
Now there is a strong analogy between this laws and the laws of thermodynamics. In fact if we look at (\ref{law3}) 
this is formally identical to the second law of thermodynamics. For the black holes so the horizon area should play the 
role of an entropy. Following this analogy to the laws of thermodynamics the surface gravity $\kappa$ has then the
role of a temperature and the law (\ref{law1}) is then analogous to the zeroth law of thermodynamics.
The laws (\ref{law2}) and (\ref{law4}) are then equivalent  respectively to the first and third law of thermodynamics.
This analogy is purely formal at classical level. In fact, using the definition  (\ref{blholedef}), a black hole cannot 
have a thermal radiation as  ``nothing can go out'' of it. A black hole should therefore have zero temperature
preventing so to speak of black hole thermodynamics. The things change dramatically as one considers the evolution 
of quantum matter fields near near the horizon, considering the metric as a fixed background.
Using this semiclassical approach Hawking discovered \cite{hawking} that a black hole emits a thermal radiation proportional to $\kappa$
  
\begin{equation}
T=\frac{\kappa}{2\pi} \; . \label{:radiation}
\end{equation}
So the 4 laws of black hole mechanics are not only a formal analogy to the laws of thermodynamics but are really the laws of thermodynamics applied
to black holes. So one can conclude that a black hole possesses an entropy given by the Bekenstein-Hawking formula \cite{bekenstein}, \cite{bekenstein2}

\begin{equation}
S=\frac{A}{4} \; .   \label{entropy}
\end{equation}
It remains the strange fact that the physical origin of the laws of thermodynamics is statistical, whereas the 4 laws of 
black hole mechanics are theorems of differential geometry and so exact laws. The description of microscopical states at the origin of 
thermodynamic laws is intrinsically quantum mechanical. In the case of black hole physics it  seems therefore  that quantum properties 
can be seen already at classical level.
Now in thermodynamics the entropy has a statistical interpretation in terms of microstates given by the Boltzmann formula
\begin{equation}
S=k_B \log N \; .         \label{bolzmann}
\end{equation}
So  there arises the question which are the microstates of the black hole responsible for the entropy.
As noticed in the introduction, if we consider a black hole as a collapsed star, the degrees of freedom of matter that 
formed the star are absolutely not enough to obtain the Bekenstein-Hawking entropy as this is much larger than the entropy of
a star of the same mass. Therefore the relevant degrees of freedom responsible for the black hole entropy are
the gravitational degrees of freedom. To describe gravitational microstates one would need a quantum theory of 
gravity. Now unfortunately for the moment there exists no complete quantum theory of gravity. The existing 
approaches to quantization of gravity like superstrings or loop quantum gravity allow to compute the entropy of some classes 
of black holes. This approaches also if very different give always as result the Bekenstein-Hawking entropy and so there
seems to be a sort of ``universality principle'' regulating the black hole entropy.
One therefore argues that, as the laws of thermodynamics can be seen already at classical level, also the gravitational
microstates responsible for the black hole entropy are regulated by some classical principle. Therefore it should be 
possible to ``count'' the microstates of a black hole without any assumption on a quantum theory of gravity.
It is important to notice, that if we find at classical level a symmetry principle, then also the possible quantum 
theory should inherit this structure, perhaps with quantum corrections. Now symmetries determine many properties 
of a system but usually not the density of states. But there is one case of symmetry where exactly this happens:
it is the 2-D conformal symmetry, which will be discussed in the next section.

\section{2-D conformal symmetry}
Let us now study a symmetry group, which is enough powerful to determinate the density of states.
First of all let us define a Weyl transformation as a transformation of the form

\begin{equation}
g_{\mu \nu} \rightarrow \omega(x) g_{\mu \nu}                           \label{weyl}
\end{equation}
or in infinitesimal form

\begin{equation}
g_{\mu \nu} \rightarrow g_{\mu \nu} + \delta g_{\mu \nu} = g_{\mu \nu} + \omega(x) g_{\mu \nu}    \; .            \label{weylinf}
\end{equation}
An action that is invariant under a Weyl transformation is called Weyl invariant. Let us now check the variation of an action under a 
Weyl transformation in arbitrary spacetime dimension using (\ref{weylinf})

\begin{equation}
\delta S = \int d^n x \frac{\delta S}{\delta g_{\mu \nu}} \delta g_{\mu \nu} = \int d^n x T^{\mu \nu} \omega (x) g_{\mu \nu} = \int d^n x \omega(x)\, T^{\nu} _{\nu}
                \; .                   \label{traceless}
\end{equation}
We can therefore conclude from (\ref{traceless}), that, in order to have Weyl invariance, the stress energy tensor must be traceless
\begin{equation}
T^{\nu} _{\nu} =0  \; .
\end{equation}
Now a coordinate transformation $ x \rightarrow x'$ that transforms the metric

\begin{equation}
g_{\mu \nu} \rightarrow g_{\mu \nu} '(x') =
 \frac{\partial x^{\rho} }{\partial { x'}^{\mu}  } \frac{\partial x^{\sigma} }{\partial {x'}^{\nu}  } g_{\rho \sigma} (x)    \label{metrictransf}
\end{equation}
in such a way that it is a Weyl transformation (\ref{weyl}) is called conformal transformation. As a special case let us take 
the metric $g_{\mu \nu}$ to be the Minkowski metric and the scale factor $\omega(x)$ to be $1$. This would then be the definition of the Lorentz
transformation. Therefore the Lorentz transformation is a special case of conformal transformation.\\
Let us now take as example the action of a free scalar field in arbitrary dimension and see if it is conformally invariant

\begin{equation}
S = \int d^n x \partial _{\mu} \phi \partial _{\nu} \phi g^{\mu \nu} \; .              \label{freeaction}
\end{equation}  
Now the stress energy tensor is
 
\begin{equation}
T_{\mu \nu} = \frac{\delta S}{\delta g^{\mu \nu}} =
 \partial _{\mu} \phi \partial _{\mu} \phi - \frac{1}{2} g_{\mu \nu} g^{\rho \sigma} \partial _{\rho} \phi \partial _{\sigma} \phi
                                                          \label{freestress}
\end{equation}
and the trace therefore is

\begin{equation}
T^{\nu} _{\nu} = \partial_{\nu} \phi \partial ^{\nu} \phi - \frac{1}{2} d \partial _{\nu} \phi \partial ^{\nu} \phi \sim \left( 1-\frac{1}{2}d  \right)
               \; ,                  \label{freetrace}
\end{equation}
where $d$ is the spacetime dimension. So we see that only in the case $d=2$ the trace of the stress energy tensor is zero.
therefore only in two dimensions the free scalar field action is conformally invariant.
Generally speaking the conformal group in two dimensions is much more powerful than in any other dimensions.
To understand why let us introduce  complex coordinates for a Euclidean flat metric
\begin{equation}
z = x^1 + ix^2      \; \; \; \; \;   \overline{z} = x^1 - ix^2 \; .
\end{equation}
The flat metric can then  be written in the form
 
\begin{equation}
ds^2= dzd\overline{z}                   \label{complexflat}
\end{equation}
and we see that every analytic and antianalytic  function 

\begin{equation}
z\rightarrow f(z) \; \; ; \; \;\overline{z} \rightarrow \overline{f}(\overline{z})             \label{analyticfunction}
\end{equation}
define a conformal transformation. In fact the metric under (\ref{analyticfunction}) transforms as

\begin{equation}
ds^2 = \frac{\partial f}{\partial z} \frac{\partial \overline{f}}{\partial \overline{z}}=\rho (z\overline{z})dzd\overline{z} \; .     \label{analyticmetric}
\end{equation}
So  the conformal group in two dimensions is infinite dimensional and therefore much larger than in any other dimension where it is finite
dimensional. The infinitesimal generators of  the transformations (\ref{analyticfunction}) are 

\begin{equation}
G_n=z^{n+1} \partial _z                  \label{infgenerator}
\end{equation}
 and they close a close a lie algebra

\begin{equation}
[G_n , G_m] =(n-m)G_{n+m}  \; .                    \label{lie}
\end{equation}
Let us now write the stress energy tensor in complex coordinates for flat Euclidean metric. It is easy to obtain
 
\begin{equation}
T_{z \overline{z}} = T_{\overline{z} z} = \frac{1}{4} ( T_{11} + T_{22} ) = \frac{1}{4}T_{\mu} ^{\mu}        \label{nondiag} 
\end{equation}

\begin{equation}
T_{zz}= \frac{1}{4} (T_{11} - 2iT_{21} - T_{22})   \; \; \; ;  \; \; \; T_{\overline{z} \overline{z}} = \frac{1}{4}(T_{11} + 2iT{21} -T_{22})
\; .                 \label{diagstress}
\end{equation}
Now the tracelessness condition (\ref{traceless}) implies that the components $T_{z \overline{z}}$ and $T_{\overline{z} z}$ in (\ref{nondiag})
are zero and therefore in 2D  the stress energy tensor has only two components 
\begin{equation}
T= T_{zz} \; \; ;  \; \;\overline{T}= \overline{T}_{\overline{zz}}   \; .                 \label{complexstress} 
\end{equation}
Using now the usual conservation law
\begin{equation}
\partial ^{\mu} T_{\mu \nu} =0
\end{equation}
we obtain in complex coordinates two relations
\begin{equation}
\partial _{z} T_{\overline{z} \overline{z}} +\partial_{\overline{z}} T_{z \overline{z}} =0 \; \; \; ;  \; \; \;
\partial _{\overline{z}} T_{zz} + \partial _z T_{\overline{z} z} =0 \; .            \label{conservation}   
\end{equation}
Using now again the tracelessness condition $T_{z \overline{z}} = T_{\overline{z} z} =0$ the previous equation becomes
\begin{equation}
\partial _z \overline{T} =0 \; \; ; \; \; \partial _{\overline{z}} T =0                \label{analyticstress}
\end{equation}
Therefore the two nonzero components of the stress energy tensor are  respectively analytic and antianalytic.
Condition (\ref{analyticstress}) of course does not prevent the components $T$ and $\overline{T}$ to have singularities.
Therefore we should use to be precise the word meromorphic instead of analytic.
A meromorphic function can be written as Laurent series. For the component $T$, for example, we obtain

\begin{equation}
T(z)=\sum _{-\infty} ^{\infty} \frac{L_n}{z^{n+2}} \; .                           \label{laurent}
\end{equation}
Now if we want to quantize to theory the stress energy tensor becomes an operator. As usual in quantum field theories 
we have to introduce normal ordering for the creation and annihilation operators in order finite quantities.
This is equivalent to subtract an infinite vacuum energy.
The result of this procedure is that the stress energy tensor does not more transform as a tensor under a
coordinate transformation $z \rightarrow w=f(z)$
\begin{equation}
T_{zz} \rightarrow \left( \frac{d w}{dz}  \right) ^2 T_{ww} + \frac{c}{12} S[w,z] \; ,           \label{anomaltensor} 
\end{equation}  
Where the quantity $S[w,z]$ is called Schwartzian derivative and is defined in (\ref{schwartz}).
The constant $c$ is called the central charge. If the constant $c$ was zero it would have been the usual tensor transformation law,
but in general this constant isn't zero. For a bosonic free field in $2-D$ e.g. we have $c=1$.
This anomaly in the transformation law arises because the renormalization breakes scale invariance.
In the quantum case also the $L_n$ become operators and because of (\ref{anomaltensor}) relatively to the commutator
they don't close the conformal algebra  (\ref{lie}) but form a Virasoro algebra \cite{polyakov}
\begin{equation}
[L_n , L_m]= (n-m)L_{n+m} + \frac{c}{12}(m^3 -m)\delta _{m, -n} \; .  \label{virasoro1}
\end{equation}
This is a central extension of the algebra (\ref{lie}). Because on the right side we have an operator the central charge must be 
seen as an operator which commutes with every element of the algebra.\\
Let us now see why the $2-D$ conformal symmetry should be relevant for the microstate counting of black holes.
It is a standard result for quantum CFT in 2-D that the asymptotic density of states for given $L_0$ is completely determined by
the Virasoro algebra by means of the  Cardy formula \cite{cardy} (a deduction  of this formula is also given in \cite{carlipBTZ})
\begin{equation}
\rho(L_0 ) =\exp \left( 2\pi \sqrt{\frac{cL_0}{6}} \right) \; .  \label{cardy1} 
\end{equation}
Therefore for given $L_0$ and knowing the value of the central charge $c$ we know also the state density
The entropy can therefore be calculated by using the logarithm of the Cardy formula
\begin{equation}
S= \ln \rho \; .
\end{equation}
Now we have found a central extension of the conformal algebra due to a quantum effect, but in general central extensions may already appear
at classical level as noticed in \cite{arnold}. An example of a theory with a classical central charge is the Liouville Theory \cite{jackiw},
which will be shown in the next section.

\section{ Liouville theory}

The Liouville theory is the theory of surfaces with constant negative curvature.
In order to find the most general action that gives as equation of motion the equation of constant negative curvature surfaces,
let us first notice that in 2 dimensions we can always write the metric in conformally flat form

\begin{equation}
\tilde{g} _{ab} = e^{\gamma \phi} \gamma _{ij}    \; ,     \label{conformalflat}
\end{equation}                                                                                          
where $\gamma_{ij}$ is the Minkowski metric not to be confused with $\gamma$ in the exponent, which is a parameter.
Using this fact, we can more generally parametrize the 2-D metric as 

\begin{equation}
g_{ij} =  e^{\gamma \phi} \hat{g}_{ij} \; ,                 \label{refmetric}
\end{equation} 
where $\hat{g}_{ij}$ is an arbitrary non-dynamic ``reference metric''. We can conclude that in 2-D the theory of its metric is
equivalent to the theory of the field $\phi$, which is related to the metric by means of equation (\ref{refmetric}),
in the sense that all the information of the metric is encoded in the conformal factor.\\
Let us now consider the following action \cite{ginsparg}

\begin{equation}
I_{Liouv} = \frac{1}{8\pi} \int d^2 x \sqrt{\hat{g}}\left[ (\hat{\bigtriangledown} \phi)^2 + QR[\hat{g}]\phi +\frac{\mu}{\gamma ^2}
e^{\gamma \phi} \right]   \; .                           \label{liouvaction}
\end{equation}  
It is  easy to proof that this action is only Weyl invariant if the constant $Q$ is set to $Q=\frac{2}{\gamma}$ and 
together with the Weyl transformation

\begin{equation}
\hat{g} \rightarrow e^{\omega}\hat{g}  \; ,
\end{equation}
we have to shift the field $\phi$

\begin{equation}
 \gamma \phi \rightarrow \gamma \phi - \omega        \label{shift}
\end{equation}
The reason  that together with the Weyl transformation we have to shift the field $\phi $, is that this field is a piece of a metric and therefore 
transforms in a  more complicate way than scalar. In fact the metric (\ref{refmetric}) under a Weyl transformation transforms as
\begin{equation}
g_{ij} =  e^{\gamma \phi} \hat{g}_{ij} \rightarrow e^{\gamma \phi +\omega} \hat{g}_{ij}
\end{equation} 
and therefore we have the shift (\ref{shift}) for the field.
The variation of (\ref{liouvaction}) with respect to $\phi$ gives the equation of motion

\begin{equation}
-2\bigtriangledown ^2 \phi +QR[\hat{g}] + \frac{\mu}{\gamma} e^{\gamma \phi} =0    \;  ,         \label{liouvmotiongeneral}
\end{equation}
which can be written as

\begin{equation}
\left( -2\bigtriangledown ^2 \phi +QR[\hat{g}] \right) e^{-\gamma \phi} =- \frac{\mu}{\gamma}     \; .          \label{liouvmotiongeneral2}
\end{equation}
Now being interested in conformal field theories let us set $Q=\frac{2}{\gamma}$ and use a standard property of the 2-D scalar curvature i.e.
 
\begin{equation}
R\left[ e^{\gamma \phi} \hat{g}  \right] = e^{-\gamma \phi} \left(R[\hat{g} ] - \hat{\bigtriangledown} ^2 \gamma  \phi  \right) \; ,   \label{curvatureconform}
\end{equation}
The equation (\ref{liouvmotiongeneral2}) becomes the equation of surfaces of constant negative curvature

\begin{equation}
R[g] =-\frac{\mu}{2} \; ,                    \label{constcurv}
\end{equation}
where the metric $g$ is defined by equation (\ref{refmetric}). Now choosing the reference metric $\hat{g}$ to be the flat metric 
the scalar curvature becomes zero and the equation of motion (\ref{liouvmotiongeneral2}) becomes 

\begin{equation}
\bigtriangledown ^2 \phi = \frac{\mu}{2 \gamma} e^{\gamma \phi}   \; .              \label{liouvilleequation}
\end{equation}
This is the standard form of the Liouville equation as introduced by Liouville.\\
The stress energy tensor derived from the action (\ref{liouvaction}), using the conformal flat gauge after making the functional 
derivative $\frac{\delta I _{Liouv}}{\delta \hat{g}}$ becomes 

\begin{equation}
T_{ab} = \partial _a \phi \partial _b \phi - \hat{g} _{ab} \left( \frac{1}{2} \partial _a \phi \partial ^a \phi + \frac{\mu}{ 2 \gamma ^2} e^{\gamma \phi}  \right)
          + Q \left[ \hat{g} _{ab} \Box \phi - \partial _a \partial _b \phi  \right]
                                                                                     \; .              \label{liouvstress}
\end{equation}
The last term $Q[....]$ comes from the variation of the $\sqrt{\hat{g}}Q R \phi$ term in the (\ref{liouvaction}) action with respect to
the metric. It is interesting to notice that this term remains in the stress energy tensor also if we choose then      
the  conformal flat gauge (\ref{conformalflat}), where the scalar curvature is of course zero.\\
The trace of the stress energy tensor becomes , taking $Q = \frac{2}{\gamma}$ as required for conformal invariance,

\begin{equation}
T_a ^a = \frac{2}{\gamma} \Box \phi - \frac{\mu}{ \gamma ^2} e^{\gamma \phi}      \; .              \label{liouvtrace}
\end{equation}
Now using the equation of motion (\ref{liouvilleequation}) the trace becomes

\begin{equation}
T_a ^a = 0  \; .               \label{tracezero}  
\end{equation}
We obtain therefore  on shell zero trace as required for conformal field theories. 
Let us now  write the stress energy tensor in light coordinates, it's nonvanishing on shell components are 

\begin{equation}
T_{\pm \pm } = \partial _{\pm} \phi \partial _{\pm} \phi - \frac{2}{\gamma} \partial ^2 _{\pm} \phi \; .              \label{liouvlightstress}
\end{equation}
Now we can define the charges
 
\begin{equation}
Q^{\pm} _f  = \int dx^{\pm} f^{\pm} (x^{\pm}) T_{\pm \pm}      \; ,                      \label{liouvcharges}
\end{equation}
where the $f^{\pm}(x^{\pm})$ are some smearing functions. We can now define the total charge as
 $Q= Q_+ + Q_-$ and compute the Poisson brackets of the total charges obtaining

\begin{equation}
\left\{ Q_f , Q_g    \right\} = Q_{[f,g]} + \frac{1}{\gamma ^2 } \Delta (f,g) \; ,           \label{liouvpoisson} 
\end{equation}
where the square bracket $[f,g] $ is the usual Lie bracket and $\Delta $ is defined as 

\begin{equation}
\Delta (f,g) = \int \left[ dx^- f^- \partial ^3 _- g^- - g^-\partial^3 _- f^- \right] +\int dx^+ \left[ f^+ \partial ^3 _+ g^+ - 
               g^+ \partial^3 _+ f^+ \right]       \; .                             \label{centralext}
\end{equation}
So the Poisson algebra of the charges gives  central extension of the conformal algebra already at \emph{classical} level, with a central charge
proportional to $Q^2$. If we quantize the theory the central charge will then get also a quantum contribution. 
The origin of this classical nonzero  central charge is that the field $\phi$ behaves not as a tensor under conformal transformations, as we have seen
in (\ref{shift}). This fact has as consequence that also the stress energy tensor does not transform as a tensor under conformal transformations.
Introducing in fact complex coordinates $z$ and $\overline{z}$, for every conformal transformation

\[
z \rightarrow w =f(z) \; ,
\]
the stress energy tensor transforms in a similar anomalous way as already seen for the quantum case (\ref{anomaltensor}) 

\begin{equation}
T_{zz} \rightarrow \left( \frac{d w}{dz}  \right) ^2 T_{ww} + \frac{1}{\gamma^2} S[w,z] \; ,           \label{nontensor} 
\end{equation}

where the quantity $S[w,z]$ is called ``Schwartzian derivative'', already cited in the previous paragraph, and is defined as 

\begin{equation}
S[w,z] = \frac{w'''}{w'} - \frac{3}{2} \left( \frac{w''}{w'}  \right) ^2     \; .                       \label{schwartz} 
\end{equation}
We have seen therefore that the Liouville theory is a classical conformal field theory that possesses a classical central charge.
This makes this theory to an interesting candidate for our search of a CFT, that enables to count the black hole microstates.

\section{Central charge and black holes}

We have now seen that the $2-D$ symmetry is exactly the type of symmetry we searched for, namely a symmetry which is enough 
powerful to determinate the state density of a system. In fact we have seen that for  $2-D$ quantum CFT the density of states 
is determined by the Virasoro algebra (\ref{virasoro1}) by means of the Cardy formula (\ref{cardy1}).\\
The Cardy formula is valid for the quantized theory, but on the other side this formula uses only the central charge and the 
eigenvalues of the $L_0$ generator. Central charges can already arise at classical level in the Poisson algebra as seen before.
In such a situation the Virasoro algebra would be inherited by the quantum theory. The central charge acquires then a quantum 
correction, but at least at leading order the density of states can be described by the classical Virasoro algebra via Cardy formula.
In the previous section we have seen that the classical Liouville theory is an example of $2-D$ CFT with a classical
central charge. Therefore if we are able to describe a black hole by means of a $2-D$ CFT with a classical central
charge we can then count the microstates using the Cardy formula without making any assumption on quantum gravity.\\ 
Now physical black holes live in ad $4-D$ spacetime. Therefore how can we use a $2-D$ theory to describe a black hole?
To answer this question let us first of all notice that being the entropy of the black hole proportional to the area of it's horizon,
we expect the gravitational degrees of freedom responsible of the entropy to live on or near the horizon.
We are therefore interested in the form of the metric near the horizon. For our purpose therefore it is the same if the black 
hole is embedded in an asymptotically flat De Sitter or Anti De Sitter spacetime. As said already in the introduction in 
this thesis we are especially interested in Spherically symmetric black holes as in existing symmetry based entropy computations they seem
to make the largest problems.\\
Now for a spherically symmetric black hole near horizon the metric always has the form

\begin{equation}
ds^2 = -N^2 (r)dt^2 + \frac{1}{N^2 (r)} dr^2 + r^2 d\Omega ^2 \; .          \label{sphericalbh}
\end{equation} 
The function $N$ is called lapse function and has a simple zero (a double zero in the extreme case) at the horizon, whereas 
$d\Omega ^2 $ is the metric of the 2-sphere.\\
From this metric we see that all the relevant geometry is encoded in the $r-t$ plane. In fact this is the part of the metric
that contains the singularities. This fact can also be seen by the Euclidean approach to black hole thermodynamics.
Therefore there is indeed a natural 2-D submanifold given by the $r-t$ plane.\\
Having now found a 2-D submanifold describing the black hole in what way can we find a Virasoro algebra?
One way is to make a dimensional reduction of the Einstein-Hilbert action integrating away the angular degrees of freedom.
One obtains so a 2-D effective theory namely a dilatonic gravity. Using a near horizon approximation we can check if we 
obtain a Liouville theory. If this happens we can compute the classical central charge and use then the Cardy formula to compute
the black hole entropy. This method will be described in the chapter 4. \\
Another possible way inspired by the work of Brown \& Henneaux \cite{brown&henneaux} is to construct diffeomorphisms that preserve 
certain fall off conditions of the $r-t$ plane metric of the black hole and then compute the Poisson brackets of the canonical
generators of this diffeomorphisms and check if they form a Virasoro algebra.
Brown \& Henneaux showed in the article cited before that this happens in $AdS_3$ for diffeomorphism generators that preserve the form
of the metric at infinity.\\  
The Poisson bracket of the canonical generators of diffeomorphisms in fact has the form \cite{brown&henneaux}
\begin{equation}
\left\{  H[\xi ] , H[\eta ]  \right\} = H\left[ [\xi ,\eta ]_{SD}   \right] +K (\xi , \eta )   \; ,    \label{poisson}
\end{equation}
where the bracket $[\, , \, ]_{SD}$ is the so called surface deformation algebra given by
\[
\left[\hat{\xi} , \hat{\eta}   \right] _{SD} ^{\bot} = \hat{\xi} ^a \partial _a \hat{\eta} ^{\bot} - \hat{\eta} ^a \partial _a \hat{\xi} ^{\bot}
\]
\begin{equation}
\left[ \hat{\xi} , \hat{\eta} \right]_{SD} ^a = \hat{\xi} ^b \partial _b \hat{\eta} ^a -\hat{\eta} ^b \partial _b \hat{\xi} ^a 
+ h^{ab}\left( \hat{\xi} ^{\bot} \partial _b \hat{\eta} ^{\bot} - \hat{\eta} ^bot \partial _b \hat{\xi} ^{\bot}  \right)   \;,
\end{equation}
where $h^{ab}$ is the metric of the spacelike hypersurface and we have introduced for $\xi$ the components normal and tangent to the 
foliation
\[
\hat{\xi} ^{\bot} = N \xi ^t 
\]
\begin{equation}
\hat{\xi} ^a = \xi ^a + N^a \xi ^t
\end{equation}
The origin of the classical central term $K (\xi , \eta )  $ in the Poisson algebra is different then in the Liouville case.
In the Liouville case in fact the origin of a classical central charge was due to the non-tensor character of the field $\phi$ in (\ref{shift}).
Whereas in this case the origin of the central extension is due to the existence of arbitrary terms, which do not depend on 
the canonical variables in the generators. In fact the canonical generators in (\ref{poisson}) have the form
\begin{equation}
H[\xi] = \int _{Bulk} \left( \hat{\xi} ^{\bot} \mathcal{H}  + \hat{\xi} ^i \mathcal{H} _i   \right) + J[\xi ] + C(\xi ) \; .     \label{canonicalgenerator}
\end{equation} 
The terms in the bulk integral $\mathcal{H}$ and $\mathcal{H _i}$ are constraints. The Boundary terms $J[\xi ]$ are needed 
to make the generators differentiable and so to make the Poisson brackets well defined (see next chapter). The term $C(\xi )$
is an arbitrary function that does not depend on the canonical variables.
Now it has been shown \cite{brown&henneaux2} that the Poisson bracket of two differentiable generators is also differentiable
and therefore has the correct boundary term. What may happen is that the Poisson bracket does not match the arbitrary term $C(\xi )$
and this is then the origin of the central term $K$.\\
This approach has been used by Strominger \cite{strominger} to compute the entropy of the BTZ black hole using it's asymptotic $AdS$ structure
and so the results of Brown \& Henneaux. The problem is that this approach is limited to the BTZ model and it's $AdS$ structure at infinity.
An extension to black holes embedded in $AdS_2$ is given in \cite{cadoni1} \cite{cadoni2}.\\
Using symmetries at infinity one is not able to distinguish a black hole from a star. On the other hand, as explained before
 one expects the degrees of freedom responsible for the entropy to live on on or near the horizon. One should therefore study
the diffeomorphisms preserving the near horizon structure instead of spatial infinity.\\   
This approach has been performed in many articles, but there seem to be some technical difficulties.
In  e.g. \cite{carlip},  \cite{ghosh} the calculation does not work in the case of non-rotating black holes. 
In \cite{carlip2}, \cite{silva}, a covariant formalism  as developed by Wald et al. \cite{wald1} \cite{wald2} \cite{wald3} \cite{wald4} is used 
instead of the original ADM approach. A generalization of this approach to lagrangians of  arbitrary curvature dependence
is found in \cite{pallua2}.\\
In our work we prefer to use 
again the canonical ADM formalism because of it's better transparency and it's successful use in the work of Brown \& Henneaux.
A return to the ADM formalism was already tried in \cite{park2}, but the boundary conditions and so the nature of the boundary terms of the canonical
generators is not completely clear (in the sense of what exactly is held fixed on the boundary).
The problem is also that in other articles \cite{koga}, \cite{hotta&sasaki} it is shown
that the central charge should be zero. Due to this discrepancies we want to analyze again this problem in the next chapters
starting this time directly with the non-rotating case, which seems to be more difficult, and paying special attention
to the different possible boundary conditions on the horizon and the associated boundary terms of the generators.
We will see that in order to have a boundary term in the hamiltonian we need to fix the surface gravity on the horizon rather than the metric. 
The crucial point in this calculations in fact is   that in order to find the central term of (\ref{poisson})
one uses the fact that the bulk part of the generators is a sum of constraints and therefore zero on shell \cite{ADM}.
On shell therefore the Poisson algebra (\ref{poisson}) reduces to the Dirac algebra of the boundary terms.
\begin{equation}
\left\{ J[\xi ] , J[\eta ]    \right\} _D = J\left[ [\xi , \eta ]_{SD}  \right] +K(\xi , \eta )  \; . \label{dirac} 
\end{equation}
As said the role of the boundary terms is to make the generators differentiable, but  problem is, that, what boundary terms are needed,
depends on the exact boundary conditions of the problem. The different possible boundary conditions for black holes 
and the associated boundary terms will be discussed in the next chapter.\\
It is important to notice that in our different approaches of a description of the entropy by means of 2-D CFT in the next chapters  we are using only 
gravitational degrees of freedom, i.e. the metric of the $r-t$ plane. Another approach can be to study a field propagating in a 
black hole metric and to find in a near horizon approximation a conformal field theory for this field as in \cite{padma}.

\chapter{Boundary terms}

\section{Bifurcation term}

Let $\phi (x)$ be an arbitrary field and let $I$ be a functional of the form
\begin{equation}
I[\phi (x) ] = \int _M \mathcal{L} (\phi (x)) dx        \; , 
\end{equation} 
where $M$ is a region of spacetime and  $\mathcal{L}$ is a function of $phi$ and of it's derivatives
\begin{equation}
\mathcal{L} = \mathcal{L} \left( \phi  , \, \nabla \phi , \dots , \nabla ^n \phi  \right)
\end{equation}
We say that a functional of this form is differentiable if for a variation of $I$ with respect to $\phi (x)$ if  we can write
\begin{equation}
\delta I = \int _M \chi \delta \phi \; .
\end{equation}
The quantity $\chi$ is called functional derivative and symbolically we can write
\begin{equation} 
\chi = \frac{\delta S}{\delta \phi}  \; .
\end{equation}
In order to have a well defined least action principle, it is necessary to have a differentiable action.
This means especially that it's variation should consist only of a bulk term without boundary terms. Boundary terms in the 
action arise due to partial integration, where one transforms total divergences in boundary integrals.
A variational principle must also be accompanied by boundary conditions. The most usual boundary condition is to keep the 
variation of the field fixed on the boundary. The variation of the Klein Gordon action for example (\ref{freeaction})
contains a boundary term with $\delta \phi $. Using therefore the standard boundary condition of fixed field on the 
boundary this action is differentiable.\\
More problematic is the case of the Einstein-Hilbert action
\begin{equation}
I_{EH} = \frac{1}{16\pi } \int _M \sqrt{-g}\, R   \; .             \label{einsteinhilbert}
\end{equation}
The scalar curvature in fact contains second derivatives of the metric (\ref{rvariation}) and therefore the boundary term contains
variations of the normal derivatives of the metric. So the action (\ref{einsteinhilbert}) with standard boundary conditions is not 
differentiable. Therefore in order to have a well defined variation principle we must add to this action a boundary term that 
cancels the boundary term arising from the variation. The correct action for smooth boundaries is \cite{hawking&gibbons}
\begin{equation}
I = \frac{1}{16\pi } \int _M \sqrt{-g} \, R + \frac{1}{8\pi} \int _{\partial M} K \,  \sqrt{h}   \; ,     \label{einsthilboundary} 
\end{equation}
where $K$ is the extrinsic curvature of the boundary and $h$ is the determinant of the boundary metric.
When calculating the hamiltonian from the action we have to keep this boundary terms. The consequence is that also
the hamiltonian has boundary terms. The boundary terms of the hamiltonian can also be found directly making the 
variation of the bulk term of it as done for asymptotically flat spaces in \cite{regge&teitelboim}.\\
In the canonical formalism we have spacetime regions of the form $ M = [t_1 , t_2 ] \times  \Sigma$, where $\Sigma$ is a spacelike 
hypersurface. For such a region the boundary has the form
\begin{equation}
\partial M = \Sigma _1 \cup \Sigma _2 \cup B^3 \; ,           \label{canboundary}
\end{equation} 
where $\Sigma _{1,2}$ are the initial and final hypersurfaces and $B^3$ is the timelike  3-boundary spatially bounding the system.
Using this notation and with this kind of boundary  with this kind of boundary the action (\ref{einsteinhilbert}) becomes,
calling $\Theta$ the extrinsic curvature of $B^3$ and $m$ the determinant of it's metric
\begin{equation}
I_0 = \frac{1}{16\pi} \int _M \sqrt{-g} \, R - \frac{1}{8\pi} \int ^{\Sigma _2} _{\Sigma _1} K \, \sqrt{h} +\frac{1}{8\pi} \int _{B^3} \Theta \, \sqrt{m} \; ,
                                                                                           \label{boundaryaction}
\end{equation}
where the integral $\int ^{\Sigma _2} _{\Sigma _1}$  means the integral over $\Sigma _2 $ minus the integral over $\Sigma _1$.
For more notational transparency let us put the the symbols used for the metric and curvature for a spacetime with
a boundary of the form (\ref{canboundary}) in a table
\begin{tabular}{l}
\\
\begin{tabular}{|l|l|l|l|l|} \hline
                           &          &     metric     &    extrinsic curvature   &   normal unit vector  \\  \hline
spacetime                   &   $M$       &     $g_{ij}$   &                       &                        \\   \hline
spacelike hypersurfaces      &$\Sigma _t$  &    $h_{ij} $    &   $K$                 &   $u^a$                  \\ \hline
3-boundary                  & $B^3$       &     $m_{ij} $   & $\Theta$              &   $\xi ^a$             \\  \hline
joints                     &   $B_{1,2}$  &     $\sigma_{ij}$&                      &   $\tilde{\xi}^a$ and $\tilde{u}^a$  \\ \hline
\end{tabular}  
\\
\\
\end{tabular}
To be precise the action (\ref{einsthilboundary}) is correct for smooth boundaries.  With a boundary of the form (\ref{canboundary})
the intersections of $B^3$ with $\Sigma _{1,2}$ are nonsmooth. The variation of the action acquires also boundary terms for the joints \cite{hayward}.
Performing in fact the variation of the actions without boundary terms   introducing the notations 
\begin{equation}
v_{ab} \equiv \delta g_{ab} \; \; \; ; \; \; \; v \equiv g^{ab} v_{ab}   \; ,
\end{equation}
we obtain the formula \cite{wald}
\begin{equation}
\delta I_{EH} = \mathrm{bulk \, terms} +\frac{1}{16\pi} \nabla _a \left( -\nabla _b v^{ab} +\nabla ^a v\right) \equiv \mathrm{bulk \, terms}
 +\frac{1}{16\pi} \nabla_a \delta \mathcal{Z} ^a  \; .
\end{equation}
The total divergence comes from the variation $g^{ab}\delta R_{ab}$. The total divergence can be converted in a boundary integral and so discarding the 
bulk terms giving the equations of motion we can write
\begin{equation}
\delta I_{EH} = \frac{1}{16\pi} \int _{\delta M} n_a \delta\mathcal{Z} ^a \; .   \label{bulvariation}
\end{equation}
Focusing for a moment on the final and initial spacelike hypersurfaces $\Sigma _{1,2}$ we can write  $\delta \mathcal{Z} ^a$ in terms of 
$K_{ab}$ and $h_{ab}$ 
\begin{equation}
u_c \delta \mathcal{Z} ^a = -2\delta K - g^{ab} \delta K{ab} + D _a \delta u^a \; .      \label{versori0}
\end{equation} 
Considering now the variation of the boundary terms associated to $\Sigma _{1,2}$ in the action $I_0$ (\ref{boundaryaction}) and combine them
with the boundary terms of the variation of the bulk action we obtain
\begin{equation}
-\frac{1}{16\pi} \int ^{\Sigma _2} _{\Sigma _1} \sqrt{h} u_c \delta \mathcal{Z} ^c - \frac{1}{8\pi} \delta \int ^{\Sigma _2} _{\Sigma _1}\sqrt{h} K
= \int _{\Sigma _1} ^{\Sigma _2} P^{ab} \delta h _{ab} -\frac{1}{16\pi}\int ^{B_2} _{B_1} \tilde{\xi} _a \delta u^a \sqrt{\sigma}
                                     \label{pezzo1}
\end{equation}
The boundaries $B_{1,2}$ are the intersections of the hypersurfaces $\Sigma _{1,2} $ with the 3-boundary $B^3$.  
The vector $\tilde{\xi}$ is the normal to $ B_{1,2}$ as considered embedded in $\Sigma_{1,2}$,  it is  equal to the 
normal $\xi$ of $B^3$ only if the boundaries are orthogonal.\\
The procedure for the 3-boundary terms of  $B^3$ is analogous as made before for the initial and final hypersurfaces. Again we have 
\begin{equation}
\xi _a \delta \mathcal{Z} ^a = -2\delta \Theta - \Theta ^{ab} \delta m_{ab} +\tilde{D} _a \delta \xi ^a  \; .
\end{equation} 
Again we put together the variation of the boundary terms of $I_0$ and the boundary terms of the bulk variation
\[
-\frac{1}{16\pi} \int _{B^3}  \sqrt{-m} \xi_c \delta \mathcal{Z} ^c - \frac{1}{8\pi} \delta \int  _{B^3}\sqrt{-m} \Theta
\]
\begin{equation}
=- \int _{B^3} \Pi^{ab} \delta m _{ab} -\frac{1}{16\pi}\int ^{B_2} _{B_1} \tilde{u} _a \delta \xi^a \sqrt{\sigma}
                                                                      \label{pezzo2} 
\end{equation}
Here the vector $\tilde{u}$ is the normal to $B_{1,2}$ as considered embedded in $B^3$ and again it is equal to the normal $u$ only when the 
boundaries are orthogonal. \\
Now we want to put the joint pieces containing $\tilde{u} _a \delta u^a$ and $\tilde{\xi} _a \delta u^a$ together. To do this
we notice that the vectors $\tilde{u}$ and $\tilde{\xi}$ can be written as
\begin{equation}
\tilde{u} = \lambda ( u -\eta\xi ) \; \; \; ; \; \; \; \tilde{\xi} = \lambda ( \xi + \eta u )  \; ,    \label{versori1} 
\end{equation}
where $\eta$ is the scalar product $\eta \equiv u\cdot \xi$ normalization factor $\lambda$ is 
\begin{equation}
\lambda = (1 + \eta ^2 ) ^{-\frac{1}{2}} \; .                                                          \label{versori2}
\end{equation}
In order to put together the terms with  $\tilde{u} _a \delta u^a$ and $\tilde{\xi} _a \delta u^a$ we can introduce the boost parameter $\theta$
defined as
\begin{equation}
\sinh \theta = u \cdot \xi \equiv \eta \; .          \label{boostparameter}
\end{equation}
Noticing now that $u\delta u = 0 $ we can write
\begin{equation}
\tilde{\xi} \delta u = \lambda \xi \delta u = \lambda \delta \eta = \delta \theta           \label{pezzo3}
\end{equation} 
we can therefore write
\begin{equation}
\int ^{B_2} _{B_1} \tilde{\xi} _a \delta u^a \sqrt{\sigma} = \int^ {B_2} _{B_1} \delta \theta \sqrt{\sigma} \; .   \label{pezzo4}
\end{equation}
In the same way we compute the term coming from the 3-boundary
\begin{equation}
\tilde{u} \delta \xi = \lambda u\delta\xi = \lambda \delta \eta = \delta \theta          \label{pezzo6}                    
\end{equation}
and therefore gives the same contribution as the term coming from the spacelike boundary (\ref{pezzo3}). We can write
\begin{equation}
\int ^{B_2} _{B_1} \tilde{u} _a \delta \xi ^a \sqrt{\sigma} = \int^ {B_2} _{B_1} \delta \theta \sqrt{\sigma} \; .              \label{pezzo5}
\end{equation}
Putting now together the single pieces of the variation (\ref{pezzo1} , \ref{pezzo2} ) and using (\ref{pezzo4} , \ref{pezzo5} ) we obtain 
eventually for the complete variation of the action the expression
\[
\delta I_0 = \frac{1}{16\pi} \int _M G_{ij} \delta g^{\mu \nu} + \int ^{\Sigma _2} _{\Sigma_1} P^{ab} \delta h_{ab} \sqrt{h}
\]
\begin{equation} 
- \int _{B^3} \Pi ^{ab} \delta m_{ab} \sqrt{-m} - \frac{1}{8\pi} \int ^{B_2} _{B_1} \sqrt{\sigma} \delta \theta  \; ,      \label{varbolt}  
\end{equation}  
 Let us now analyze the terms on the
right side of (\ref{varbolt}). The first term gives the equations of motion the second and third are terms are linear in  to the variation
of the boundary metric and therefore with our boundary conditions zero. The last term in general is not zero.
Therefore in the presence of nonsmooth intersections of boundaries the correct action for fixed boundary metric is 
\begin{equation}
I' = \frac{1}{16\pi}\int _M \sqrt{-g} \, R - \frac{1}{8\pi} \int ^{\Sigma _2} _{\Sigma _1} K \, \sqrt{h} +\frac{1}{8\pi} \int _{B^3} \Theta \, \sqrt{m}
+\frac{1}{8\pi}\int ^{B_2} _{B_1} \sqrt{\sigma} \theta  \; .                 \label{boltaction} 
\end{equation}
The last term in the action, i.e. the joint term in literature is also called ``tilting term'' \cite{hawking&hunter}. It is zero in the
case, that the hypersurfaces $\Sigma$ are orthogonal . We have considered up to now nonsmooth boundaries given by the intersection of $\Sigma$ with $B^3$. 
Another case of nonsmooth boundary can be given by two intersecting spacelike hypersurfaces.\\
Now let us consider a static black hole. It's timelike killing vector is null on the horizon, this means that in the standard foliation $t= \mathrm{const}$ 
all the spacelike hypersurfaces intersect in a 2-D sphere called the bifurcation. Therefore in this situation the action describing a spacetime containing 
a black hole has a nonsmooth boundary in the bifurcation given by two
intersecting spacelike hypersurfaces $\Sigma _{1,2}$. This kind of joint in literature is  also called ``bolt'' \cite{hawking&gibbons2}. 
In the case of a ``bolt'' all the computation done before leading to (\ref{varbolt}) can be repeated. Using in fact again  (\ref{versori0})
we obtain now as contribution from the two spacelike hypersurfaces, converting the total divergences in an integral on the joint, the joint contribution
\begin{equation}
\Delta = \frac{1}{16\pi} \int _B  \sqrt{\sigma} \left( \tilde{\xi} _2 \cdot \delta u_1 - \tilde{\xi}_1 \cdot \delta u_2   \right)        \; .
\end{equation}  
The vector $\tilde{\xi}_2$ is the normal to the bolt as considered embedded in $\Sigma ^1$ and the vector $\tilde{\xi}_1$ is the normal to the 
bolt as considered embedded in $\Sigma _2$. Now again as in (\ref{versori1} ,\ref {versori2}) we can write the vectors $\tilde{\xi}_1$ and $\tilde{\xi}_2$
as linear combination of  $u_1$ and $u_2$ with the only change that now scalar product is $\eta \equiv u_1 \cdot u_2$.
The boost parameter $\theta$ this time is  defined as 
\begin{equation}
\cosh \theta = -  u_1 \cdot u_2     \label{boostparameter2}  \; .
\end{equation}
This is because in the bolt case, being the intersecting hypersurfaces both spacelike,  their normals cannot be orthogonal.
Following now the same procedure as in (\ref{pezzo3} , \ref{pezzo6}) it is immediate to proof that
\begin{equation}
\tilde{\xi}_2 \cdot u_1 - \tilde{\xi}_1 \cdot u_2 = -2\delta \theta  \; .
\end{equation}
The total bolt contribution from the bulk variation is therefore 
\begin{equation}
\Delta = -\frac{1}{8\pi} \int _B \sqrt{\sigma} \delta \theta
\end{equation}
Therefore if we treat the event horizon as a boundary and using the standard foliation the correct action by fixed bolt metric is   
\begin{equation}
I= I_0 +  \frac{1}{8\pi}\int _{bolt} \sqrt{\sigma} \theta  \; . \label{bifurcationaction}
\end{equation}
Let us now notice that for a Kerr black hole the killing vector $\partial _t$ on the horizon goes to zero only in two points, namely
the ``north pole'' and the ``south pole''. Having only two points in which the $t= \mathrm{const}$ hypersurfaces intersect there is no boundary term for this 
intersection. The action therefore acquires no extra term at least in the standard foliation. Having in this case only 2 points as intersection
also the hamiltonian won't have boundary terms associated to the horizon. Therefore the technique to find a central extension of the boundary terms 
Dirac algebra (\ref{dirac}) like in \cite{brown&henneaux} seems to work only for the nonrotating case, at least in the standard foliation.\\
Now we have seem that the boundary term of the action associated to the bifurcation (\ref{bifurcationaction}) works with boundary condition of 
fixed boundary metric. This is the most usual boundary condition but surely not the only possible. One for example can also fix the normal derivative 
of the  boundary metric. In the case of the bifurcation this means to fix the surface gravity.\\ 
Let us now analyze what boundary term we have to associate to the bifurcation in the case that we keep the surface gravity fixed instead of the
metric. In order to do this let us write the parameter $\theta$ of (\ref{boostparameter} for the case of a nonrotating black hole.
In this case we have metric of the form
\begin{equation}
ds^2 = -N^2 dt^2 + N^{-2} dr^2 + r^2 d\Omega ^2   \; . \label{nearhorizon}
\end{equation}
In the bifurcation all the constant $t$ hypersurfaces intersect and so the normal to the hypersurfaces there is not well defined.
to compute the scalar product in (\ref{boostparameter2}) we can parallel transport the normal of one hypersurface, say $\Sigma _{t_1}$ to another
say $\Sigma _{t_2}$ along an $r=\mathrm{const}$ curve.
Using (\ref{nearhorizon}) the only nontrivial parallel transport equations become
\begin{equation}
\dot{u}^t + \Gamma ^{t}_{tr} \dot{t} u^r =0 \; \; \; ; \; \; \; \dot{u}^r +\Gamma ^{r}_{tt} \dot{t} u^t + \Gamma ^{r}_{tr} \dot{t} u^r =0
\end{equation}
with 
\begin{equation}
\Gamma ^{t }_{tr} = \frac{1}{2} N ^{-2}  (N ^2 ) ' \; \; ; \; \;  \Gamma ^{r }_{tt} =\frac{1}{2} N ^2  (N ^2 ) ' \; \; ; \; \;  \Gamma ^{r}_{tr} =0  \; .
\end{equation}
The solution therefore is 
\begin{equation}
u^t = N^{-1} \cosh (\kappa t) \; \; \; ; \; \; \; u^r = -N \sinh{\kappa t}  \; .
\end{equation}
The scalar product of the normals of $\Sigma _1$ and $\Sigma _2$ is then 
\begin{equation}
u_1 \cdot u_2 = - \cosh (\kappa \Delta t)  \; .
\end{equation}
the boost parameter $\theta$ is then
\begin{equation}
\theta = \kappa \Delta t 
\end{equation}
this means that the bifurcation term of the action can be written as 
\begin{equation}
\frac{1}{8\pi} \int _{Bolt} \kappa dA dt \; .       \label{boltcanonical} 
\end{equation}
Now let us remember that the origin of the bolt term is to cancel the term linear in $\delta \theta$ in (\ref{varbolt}). Now being $\theta$
proportional to the surface gravity in the case we keep the surface gravity fixed in the variation the $\delta \theta$ term is then zero.
We can therefore conclude that in the case we keep the surface gravity fixed there is no bifurcation term in the action.

\section{Hamiltonians}
Now in order to find the boundary terms of the canonical generators let us write the action without tilting term (\ref{boundaryaction}) in canonical form
expressing everything in function of the canonical momenta and the hypersurface 3-metric $h_{ab}$. The result is \cite{hawking&hunter}
\[
I_0 = \int _M \left( P^{ab} \dot{h} _{ab} - N\mathcal{H} - N^i \mathcal{H} _i \right) d^4 x -2 \int dt \int _{B_t} h^{-1/2}P^{ab} N_a \tilde{\xi} _b
\sqrt{\sigma} d^2x 
\]
\begin{equation}
- \frac{1}{8\pi} \int _M \nabla _a Z^a \sqrt{-g} d^4 x -\frac{1}{8\pi} \int _{\Sigma _1} ^{\Sigma _2} K \sqrt{h} d^3 x + \frac{1}{8\pi}
\int _{B_3} \Theta \sqrt{-m} dt d^2 x      \; .              \label{hahu}
\end{equation}
The functions $\mathcal{H}$ and $\mathcal{H}_i$ are the hamiltonian constraints (see the appendix).
The boundary $B_t$ is the foliation of $B_3$ in the form
\begin{equation}
B_t = B_3 \cap \Sigma _t
\end{equation} 
The term $Z^a$ is given by
\begin{equation}
Z^a =\nabla _u u^a - u^a \nabla _b u^b
\end{equation}
and so we have that
\begin{equation}
Z^a u_a =K          \label{formula}
\end{equation}
Now as next we must convert to a boundary integral  the $\nabla _aZ^a$ term.
Notice that due to (\ref{formula}) the integral $\int _{\Sigma _1} ^{\Sigma _2}$ is therefore 
canceled. It survives only the boundary integral over $B_3$. Now in order to be able to read out the hamiltonian from (\ref{hahu})
we have  to factorize out a $\int dt$ term
from the boundary integrals. To do this let us notice that $B_3$ is foliated by $B_t$ and therefore we can write
\begin{equation}
\sqrt{-m} = N \lambda \sqrt{\sigma} \; ,
\end{equation} 
where $\sigma _{ab}$ is the metric induced on each $B_t$ by $m_{ab}$ and $h _{ab}$ and $\lambda$ is defined as $\lambda = \cosh \theta$.
The surviving boundary terms can therefore be factorized with $\int dt$
\[
\frac{1}{8\pi} \int _{B^3} \Theta \sqrt{-m} dtd^2 x -\frac{1}{8\pi} \int _{B^3} \xi _a Z^a \sqrt{-m} dt d^2 x 
\]
\begin{equation}
= \frac{1}{8\pi} \int dt \int _{B_t} N \tilde{\Theta} \sqrt{\sigma} d^2 x - \frac{1}{8\pi} \int dt \int _{B_t} N \lambda \tilde{u} ^a \partial _a (\theta)
\sqrt{\sigma} d^2 x  \; ,            \label{factorization}
\end{equation}
where $\tilde{\Theta} $ is the trace of the extrinsic curvature of $B_t$ as embedded in $\Sigma _t$ and $\tilde{u}$ is
the normal to $B_t$ as considered embedded in $B^3$. If we choose the time flow $\tau$ to be tangent to $B^3$ we can write
\begin{equation}
\tau ^a = N \lambda \tilde{u} ^a 
\end{equation}
and therefore
\begin{equation}
N\lambda \tilde{u} ^a \partial _a \theta = \dot{\theta} \; .
\end{equation}
Therefore the last term in (\ref{factorization}) becomes
\begin{equation}
-\frac{1}{8\pi} \int _{Bolt}  \theta \sqrt{\sigma} d^2 x + \frac{1}{8\pi} \int dt  \int _{B_t} \theta \dot{\sqrt{\sigma}} d^2 x      \; .
                                                                            \label{cancellation} 
\end{equation} 
The first term cancels the tilting term in (\ref{boltaction}) whereas the second term in the case that the joint is a bifurcation is zero because
it is static per definition.\\
The hamiltonian resulting from (\ref{hahu}) is using (\ref{factorization} , \ref{cancellation}) 
\begin{equation}
H = \int _{\Sigma _t} d^3 x \sqrt{h} \left( N\mathcal{H} + N^i \mathcal{H}_i \right) - \frac{1}{8\pi} \int_{B_t} 
\left( N \tilde{\Theta} - 16\pi h^{-1/2} P^{ab} N_a \tilde{\xi}_b   \right)  \sqrt{\sigma}  
\end{equation}
We see so that there is no contribution from the bifurcation to the hamiltonian boundary terms. 
The action (\ref{boltaction}) was the correct one in the case that the boundary metric was kept fixed. If we now take the case of a black hole 
bifurcation as joint with the surface gravity held fixed instead of the bolt metric we have seen that there is no tilting term contribution to the action.
Therefore there is then a tilting term in the hamiltonian that comes from (\ref{cancellation}) that is now not canceled from the action.
The hamiltonian for fixed surface gravity $H'$ is therefore
\begin{equation}
H' = H - \frac{1}{8\pi} \int _{Bolt} \theta \sqrt{\sigma} d^2x
\end{equation}
In this case there is a boundary term contribution from the bifurcation to the hamiltonian. 
We have in the case of the bifurcation therefore the situation, that fixing the metric there is a tilting term in the action but not in the 
hamiltonian. Whereas fixing the surface gravity there is a no tilting term in the action but there is one in the hamiltonian. 
Therefore if we want to have a on shell a Dirac algebra for the black hole we necessarily must fix the surface gravity.
This is physically reasonable because the surface gravity gives the temperature of the black hole and our calculations attempt 
precisely to describe the black hole thermodynamics.\\
By fixed surface gravity therefore using (\ref{boltcanonical}) the bolt term associated to the canonical generator of the vector field
$\xi$ is
\begin{equation}
J[\xi ] = - \frac{1}{8\pi} \int _{Bolt} n^c D_c \xi ^{\bot}   \label{canonicalbolt} 
\end{equation}

\section{Fall off conditions}
Now we want to study the deformations of the $r-t$ plane that preserve the surface gravity of the horizon.
In order to do this we must find the falloff condition of the vector fields generating the diffeomorphisms. Let us now find the most general expression
for a near horizon metric. As explained before we must start from the nonrotating case in order to have a bifurcation in the standard foliation.
Making the Ansatz of spherical symmetry we have
\begin{equation}
ds^2 = -N^2 (r,t) + A^2 (r)dr^2 +r^2d \Omega ^2
\end{equation}
We have to impose some conditions on the functions $N$ and $A$.\\
First of all we notice that the existence of a bifurcation implies the vanishing of the lapse function $N$ on the horizon
\begin{equation}
N^2 (r_{+},t) =0   \; .
\end{equation}
We are dealing with a system of fixed surface gravity, where the surface gravity is defined as 
\begin{equation}
\lim_{r \rightarrow r_{+}} \frac{\partial_r N}{A} = \kappa
\end{equation}
We must also impose the topology of the black hole in order to distinguish it from flat spacetime. In the Euclidean case the black hole has the topology
$R^2 \times S^2 $ and therefore the Euler characteristic is $\chi = \chi (\mathrm{disk} ) \times \chi (\mathrm{sphere})$ \cite{brownetal} and being the Euler 
characteristic of the sphere $2$, the Euler characteristic of the black hole is  
\begin{equation}
\chi = 2   \; , 
\end{equation}
whereas the flat spacetime has $\chi =0$. Calculating $\chi$ we obtain
\begin{equation}
\chi = 2 \left( 1-A^{-1} (r_{+} \right)   \; .
\end{equation}
We obtain so the condition on $A$
\begin{equation}
A^{-1}(r_{+}) =0
\end{equation}
If we impose the hamiltonian constraints we obtain the form for$A^2$
\begin{equation}
A^2 = \left(1- \frac{r_{+}}{r} \right) ^{-1}
\end{equation}
Using this conditions we obtain the explicit form for the lapse
\begin{equation}
N^2 = 4 \kappa^2 r_{+} (r-r_{+}) + a(t)(r-r_{+})^2
\end{equation}
Now on shell we have $r_{+}=\frac{1}{2 \kappa}$
therefore putting all together our diffeomorphisms must preserve the following conditions
\begin{equation}
N^2 =g_{tt}= 2 \kappa (r-r_{+}) + \mathcal{O}(r-r_{+})^2    \label{condition1}
\end{equation}
and for $g^{rr}$ we have then
\begin{equation}
g^{rr} = N^2 + \mathcal{O} (N^3 )       \label{condition2}
\end{equation} 
We must now search for vector fields which preserve the two conditions (\ref{condition1} ) and (\ref{condition2}). Satisfying this conditions the
diffeomorphisms automatically preserve the surface gravity. It may seem a contradiction, that we want to fix also the form of the near horizon metric
and not only the surface gravity, but we must consider, that in order to have a bifurcation and therefore a boundary term associated to it,
we must also preserve the existence of the bifurcation. For a vector field $\xi$ in the $r-t$ plane the variation of $g_{tt} $ is given by
\begin{equation}
\delta _{\xi} g_{tt} = \mathcal{L} _{\xi} g_{tt} = \partial _r g_{tt} \xi^r + 2g_{tt}\partial _t \xi^t = \mathcal{O} (N^3 )
\end{equation}
and therefore
\begin{equation}
\xi ^r = - \frac{N^2}{\kappa} \partial _t \xi^t + \mathcal{O}(N^3)        \label{radialdiff}
\end{equation}
The variation of the $g^{rr}$ component is now
\begin{equation}
\delta _{\xi} g^{rr} = \partial _r g^{rr} \xi ^r + 2g^{rr} \partial _r \xi ^r = \mathcal{O} (N^3)
\end{equation}
The boundary term of the generator   (\ref{canonicalbolt}) implies only the $\xi ^t$ component and therefore let us use (\ref{radialdiff}) 
in the last variation in order to find the form of $\xi ^t$
\begin{equation}
=- 2N ^2 \dot{\xi} ^t -4 N^2 \dot{\xi} ^t - \frac{2N^4}{\kappa} (\dot{\xi} ^{t})' =\mathcal{O} (N^3)
\end{equation}
This means that 
\begin{equation}
 \dot{\xi ^t }= \mathcal{O} (N) \; \; \; ; \; \; \; \xi^r = \mathcal{O}(N^3)  \; .
\end{equation}
This means that we have for the killing vector
\begin{equation}
\xi ^t =\mathcal{O} (1) \; \; \; ; \; \; \; \xi ^{\bot} = \mathcal{O} (N)   \; .
\end{equation}
Therefore with our boundary conditions we have a bondary term in the generators and therefore there is the possibility
of the existence of a nontrivial central charge. We must therefore compute the Poisson brackets for two generators.
The tedious calculation with our boundary condition can be done using the results in \cite{brown&lau&york} where the variation of the hamiltonian
under a quasilocal boost is done. The boundary terms coming from this variation can be computed going on shell obtaining eventually for
our boundary conditions
\begin{equation}
\left\{ H[\xi ] , H[\eta ]    \right\}_{PB} = H[\xi , \eta ] _{SD} \; ,
\end{equation} 
where $H$ is the generator with the correct boundary term. We conclude therefore that there is no central extension of the algebra in 
this case also if the boundary terms are nonzero.\\
We have until now analyzed two cases: the case of fixed bolt metric and the case of fixed surface gravity.
In the first case there were no boundary terms and so becoming the generator algebra the constraint algebra it does not admit a central extension.
In the second case there were nozero boundary terms associated to the bifurcation, but the computation of the Poisson brackets shows that there is no
central extension in the generator algebra.
The calculation of Strominger \cite{strominger} seems therefore limited to the $BTZ$ black hole. The $BTZ$ black hole is in fact 
embedded in an $AdS$ spacetime. The fact that one finds a nonzero central charge 
for the generators, that preserve the $AdS$ structure at infinity becomes a particular case of the  $AdS/CFT$ correspondence \cite{maldacena}.
In this case the relevant CFT at spatial infinity is the Liouville theory \cite{henneauxL}. The quantum version of this Liouville theory living on the $Ads$
boundary is described in \cite{chen}, where  again the BTZ entropy is obtained.
There is also a $dS /CFT$ correspondence \cite{strominger2}, \cite{klemm1}, \cite{klemm2}, \cite{klemm3} but we want to describe a black hole
independently of its embedding.\\   
Also if up to now we found a negative result in the next chapter we will see how to count the black hole microstates by means of the Cardy 
formula using a classical central charge. What changes then is the origin of the central charge, which will not arise from a Dirac algebra of boundary terms
but from the non-scalar transformation property of a Liouville field.

\chapter{Black holes and Liouville Theory}

\section{Dimensional reduction}

We have up to now seen that it is not possible to obtain a central charge from the Dirac algebra of the boundary terms of the canonical generators,
because there are no boundary terms associated to the bifurcation when the surface gravity is fixed. This does not mean that it is not possible
to find a Virasoro algebra that describes the black hole with some other approach.\\
Instead of studying the deformations of the $r-t$ plane in a 4-D spacetime let us use instead an effective 2-D theory that describes the black hole. 
In order to do this let us again take a spherically symmetric metric
\begin{equation}
ds^2= g^{(2)}_{ij}dx^i dx^j + \Phi^2 (x_1 ,  x_2) \left( d\theta^2  + \sin ^2 \theta d\phi ^2  \right) \; ,        \label{spheric1}
\end{equation}
where $g^{(2)}$ is the metric of  a 2-D manifold and $\Phi$ is a radial coordinate. As we have already seen this 2-D metric describes the geometry of the
black hole. Now in order to obtain an effective two dimensional theory let us take the 4-D Einstein Hilbert action
\begin{equation}
I= \frac{1}{16\pi}\int \sqrt{-g} R d^4 x  \; . 
\end{equation}
We can write the action  in terms of the 2-D scalar curvature $R^{(2)}$ and $\Phi$ and then integrate over the angular variables obtaining \cite{oneill}

\begin{equation}
I=\frac{1}{4} \int  d^2 x \sqrt{-g^{(2)}}\left( 2(\nabla \Phi)^2 +\Phi ^2 R^{(2)} +2  \right)  \; .       \label{reduced1}
\end{equation}
This is a 2-D gravity coupled to a dilaton field. This effective action describing the metric of the form \ref{spheric1} contains two fields,
namely the dilaton $\Phi$ which describes the radius of the horizon and the metric $g^{2}$ which describes the geometry of the $r-t$ plane.
In literature this action has already been used in order to obtain a conformal field theory \cite{solodukhin} \cite{brustein} describing a black
hole, focusing on the dynamics of the  dilaton field. There exists also a generalization to Gauss-Bonnet gravity with this approach \cite{pallua1}.
 Such approaches therefore interpreted the degrees of freedom responsible for the black hole entropy as 
fluctuations of the horizon radius. Our approach is different. Considering in fact that the geometry of the black hole is described by the 
metric $g^{2}$ we argue that this field should be relevant for the entropy of the black hole rather than the dilaton field. The idea is therefore 
to use a near horizon approximation which enables us to rule out the $\Phi$ field leaving us with a theory of the 2-D metric alone and check if
this theory is a conformal field theory and with a classical central charge.
In order to do this let us first of all write the action in more useful form. we redefine our two fields as 

\begin{equation}
\Phi^2 = \eta \; \; \; ;  \; \; g_{ab} ^{(2)} = \frac{1}{\sqrt{\eta}} \tilde{g}_{ab}     \label{redefine1}
\end{equation}
obtaining so the action of a dilatonic two dimensional theory in the usual form

\begin{equation}
I=\frac{1}{2} \int \sqrt{-\tilde{g}} \left[ \frac{\eta}{2} R[\tilde{g}] + V(\eta )   \right] \; ,  \label{reduced2}
\end{equation}
where the dilatonic potential $ V(\eta )$ in our case is
\begin{equation}
V(\eta ) = \frac{1}{\sqrt{\eta}}    \; .
\end{equation}
Let us study the equations of motion of the action (\ref{reduced2}). We obtain for the variation of the action  with respect to $\eta$
\begin{equation}
R[\tilde{g} ] +2\partial _{\eta} V(\eta) =0    \label{motion1}
\end{equation}
and for the variation with respect to $\tilde{g}$

\begin{equation}
\nabla _a \partial _b \eta - \tilde{g}_{ab} \Box _{\tilde{g}} \eta + \tilde{g} _{ab} V(\eta) = 0 \; .        \label{motion2}
\end{equation} 
We have already noticed that in two dimensions the metric can always be put in conformally flat form. Let us therefore write the 
field $\tilde{g}$ as

\begin{equation}
\tilde{g} _{ab} = e^{-2\rho} \gamma _{ab} \; ,                     \label{flatgauge}
\end{equation}
where $\gamma{ij}$ is as usual the Minkowski metric. We see therefore that the geometry of the $r-t$ plane is completely described by the 
Liouville field $\rho$. Using (\ref{flatgauge}) the action (\ref{reduced2}) becomes

\begin{equation}
I= \frac{1}{2} \int d^2 x \left( -\partial _a \eta \partial ^a \rho + V(\eta ) e^{-2\rho}   \right)  \; .     \label{reduced3}
\end{equation}
This is a theory of two fields propagating in flat spacetime. From this action we obtain two coupled equations of motion
\begin{equation}
\Box \rho + \partial _{\eta} V(\eta ) e^{-2\rho} =0          \label{motion3}
\end{equation}

\begin{equation}
\Box \eta - 2V(\eta ) e^{-2\rho}=0  \; .      \label{motion4}
\end{equation} 
Having now gauge fixed the metric to the form (\ref{flatgauge}) the equations of motion (\ref{motion3}) and (\ref{motion4}) must be 
implemented by the constraint

\begin{equation}
\frac{\delta I}{\delta g^{ab}} =T_{ab}=0 \; .      \label{constraint1}
\end{equation}
The stress tensor of (\ref{reduced2}) using the gauge (\ref{flatgauge}) becomes
   
\begin{equation}
2T_{ab} = -\partial _a \eta \partial _a \rho + \frac{1}{2} \partial _c \eta \partial ^c \rho \, \gamma_{ab} 
- \frac{V(\eta )}{2} e^{-2\rho} \, \gamma_{ab} - \frac{\partial _a \partial _b \eta }{2} +\frac{\gamma _{ab} \Box \eta}{2}  \; .       \label{constraint2}
\end{equation}
The equations of motion and the constraints get a very simple form introducing light coordinates $x^{\pm} = x^1 \pm x^2$, i.e. for the equations of motion
\begin{equation}
\partial _{+} \partial _{-} \rho - \frac{\partial_{\eta} V(\eta )}{4} e^{-2\rho}=0                         \label{motion5}
\end{equation}
\begin{equation} 
\partial _+ \partial _- \eta + \frac{V(\eta )}{2} e^{-2\rho} = 0                                  \label{motion6}
\end{equation}
and for the two constraints $T_{\pm \pm} =T_{11} + T_{22} \pm T_{12} =0$
\begin{equation}
T_{\pm \pm } =\partial _{\pm} \partial _{\pm} \eta + 2 \partial _{\pm} \rho \partial _{\pm} \eta =0 \; .    \label{constraint3}
\end{equation}
Therefore in light coordinates the constraints do not depend on the dilatonic potential.
It is interesting to notice that the two equations of motion (\ref{motion5}) and (\ref{motion6} are not independent.
In fact differentiating (\ref{motion6}) with respect to $\partial _-$ we obtain
\begin{equation}
\partial _{+} \left( \partial _- \partial _- \eta  \right) +\frac{1}{2}\partial _{\eta} V(\eta ) \partial _- \eta e^{-2\rho}
+\frac{V(\eta )}{2} \left( -2\partial _- \rho e^{-2\rho}  \right) =0   \; .
\end{equation}
Using the constraint $T_{--} =0$ we get 
\[
\partial _+ \left[-2\partial _- \rho \partial _- \eta \right] + \frac{1}{2} \partial _{\eta} V(\eta ) \partial _ - \eta e^{-2\rho}
-V(\eta ) \partial _- \rho e^{-2\rho}
\]
\begin{equation}
= -2 \partial _- \partial _+ \rho \partial _- \eta - 2 \partial _-
 \rho \partial _+ \partial _- \eta +\frac{1}{2} \partial _{\eta} V(\eta ) \partial _ - \eta e^{-2\rho}
-V(\eta ) \partial _- \rho e^{-2\rho} = 0     \;,
\end{equation}  
obtaining eventually
\begin{equation}
-2 \partial _- \rho \left[ \partial _+ \partial _- \eta + \frac{1}{2} V(\eta ) e^{-2\rho}    \right] - 
2 \partial _- \eta \left[ \partial _+ \partial _- \rho - \frac{1}{4}\partial _{\eta} V(\eta ) e^{-2\rho}  \right] =0   \; .    \label{equivalence}
\end{equation}
The first term in the last equation is simply eq. (\ref{motion6}) which is proportional to eq. (\ref{motion6}) and therefore zero.
The second term in is proportional to eq.(\ref {motion5}). We can therefore conclude that eq. (\ref{motion6}) plus the 
constraints imply eq. (\ref{motion5})  
This suggests that the solution of the equations of motion, which involve two fields actually may be written in terms of a single field.

\section{Near horizon approximation}

Up to now we have only made the Ansatz of spherical symmetry. In this section we want to analyze the effects of the presence of a black hole 
on the equations of motion. We are, as in the previous chapters, in the near horizon behavior of our fields. As seen before the metric of the
$r-t$ plane of a spacetime containing a black hole near horizon can be approximated by

\begin{equation}
ds^2 = -N^2 (r)dt^2 + \frac{1}{N^2 (r)} dr^2 \; ,        \label{bhmetric1}
\end{equation}
where the lapse function $N$ near horizon at the first order has the form
 \begin{equation}
N^2 (r) = 2\kappa (r-r_0 )   \; .    \label {lapse1}
\end{equation}
We want to find a near horizon solution for the conformal factor $\rho$ in order to do this let us introduce the coordinate
\begin{equation}
r = r_0 + \frac{\kappa y^2}{2}
\end{equation}
so that the metric (\ref{bhmetric1}) becomes

\begin{equation}
ds^2 = -\kappa ^2 y^2 dt^2 + dy^2    \;.      \label{rindler1}
\end{equation}
Now introducing tortoise coordinates 
\begin{equation}
x^{\pm} = \kappa t \pm \log y
\end{equation} 
we eventually obtain a conformally flat metric

\begin{equation}
ds^2 = -dx^+ dx^- \exp \left(  x^+ - x^- \right) \; .          \label{Giacomini:rindler2}
\end{equation}
Comparing this expression with the definition of the conformal factor $\rho$ (\ref{flatgauge}) we can write
the form of the conformal factor near the horizon as 
\begin{equation}
-2\rho = x^+ - x^-     \; .      \label{factor1}
\end{equation}
Let us now notice that the constraints (\ref{constraint3}) can be written as 
\begin{equation}
\frac{ \partial _{\pm} \left( \partial _{\pm} \eta  \right)}{\partial _{\pm} \eta} = -2 \partial _{\pm} \rho
\end{equation}
\begin{equation}
\partial _{\pm} \ln \left( \partial _{\pm} \eta  \right) = -2 \partial _{\pm} \rho   \; .
\end{equation}
It is therefore easy now to integrate the constraints 
\begin{equation}
\ln \partial _{\pm} \left( \eta \right) = -2\rho + C_{\mp} (x^{\mp})
\end{equation}
obtaining eventually
\begin{equation}
\partial_{\pm} \eta = \exp \left( -2\rho + C_{\mp}(x_{\mp}  \right) \; ,      \label{integrated1}
\end{equation}
Where the function $C_{\mp}$ is an arbitrary function of the coordinate $x^{\pm}$. If we take now the limit of approaching the horizon
i.e. taking $x^{\pm} \rightarrow \mp \infty$, which because of eq. (\ref{factor1}) means $\rho \rightarrow \infty$ the integrated constraint 
reduces to

\begin{equation}
\partial_{\pm} \eta = 0 \Rightarrow \eta = \mathrm{const}  \; .                        \label{constantdilaton}
\end{equation}  
This means that in the near horizon limit the dilaton field can be considered as fixed at it's value on the horizon $\eta_0$.
In that case it is immediate to check that the constraints (\ref{constraint3}) and the  equation of motion  
(\ref{motion6}) are identically satisfied. As equation of motion survives therefore only eq. (\ref{motion5}), which now takes the form
\begin{equation}
\partial _+ \partial_- \rho  + \frac{1}{8\eta _0 ^{3/2}} e^{-2\rho}  =0   \; .           \label{liouville1}
\end{equation}
This is the already seen Liouville equation (\ref{liouvilleequation}).\\
We have started with a metric with spherical symmetry and found that the dynamic of this metric is regulated by two coupled field equations
(\ref{motion5}), (\ref{motion6}). Using a near horizon approximation we have found that the equation for the conformal factor field decouples 
becoming a Liouville equation and the dilaton field is ruled out. Therefore as already argued all the dynamics of the black hole is regulated 
only by the conformal factor $\rho$ which describes the $r-t$ plane geometry. We have already seen that the Liouville Theory possesses a 
classical central charge. It is not a surprise that this is true also for the theory of the conformal factor $\rho$, in fact this field 
is a part of a metric and therefore transforms not as a scalar. This was also the origin of the anomaly in the Liouville theory.
In the next section we want to construct explicitly the Virasoro algebra and compute it's central charge.

\section{Virasoro algebra}
Having now found that near horizon we have a Liouville theory let us consider the the most general action for the field $\rho$ that gives the Liouville
equation. It's form is given by (\ref{liouvaction}) 
\begin{equation}
I = C\int \sqrt{\hat{g}} \left( \frac{1}{2} \partial_{\mu} \rho \partial^{\mu} \rho + \frac{\mu}{\beta ^2}e^{-\beta \rho} - 
\frac{2}{\beta} \rho R[\hat{g}]  \right)    \; .             
\end{equation}
In our case $\beta =2$ and $\mu =-2 V' (\eta _0)= \frac{1}{\eta _0 ^3/2}$. The constant $C$ for the moment is not determinated. In fact we 
have not derived this action from dimensional reduction but we taken it as the action that leads to the equation (\ref{liouville1}).
In our case we have chosen the reference metric $\hat{g}$ to be the Minkowski metric $\gamma$ therefore the scalar curvature $R$ vanishes.
But as seen in the section on Liouville theory the scalar curvature gives a contribution to the stress energy tensor also if we take the 
conformal flat gauge. Now in order to define Virasoro generators we must again consider the stress tensor of the Liouville action
(\ref{liouvlightstress}). In our case it becomes 
\begin{equation}
T_{\pm \pm} = C \left( \partial _{\pm} \rho \partial _{\pm} \rho + \partial _{\pm} \partial_{\pm} \rho  \right) \; . 
                                                                      \label{liouvlightstress2}
\end{equation} 
The term with the second derivatives in (\ref{liouvlightstress2}) comes again from the $R \rho$ coupling and is therefore 
analogous to the (\ref{varcoupling3}) term in standard coordinates. The total central charge associated to this theory is given by (\ref{liouvpoisson})
\begin{equation}
c = 12 C \frac{4}{\beta ^2}  \; .
\end{equation}
The central charge and the stress tensor  depend therefore from the constant $C$ but neither from the dilaton value $\eta _0 $ nor from $\mu$.
In order to continue we must now fix the constant $C$ in the action. We fix the constant in such a way that the energy of the system equals
the energy of the black hole i.e. it's ADM mass.In fact it sounds natural that an effective theory describing that  black hole should 
have the same energy.  In order to compute the energy of the system we use the near horizon solution for $\rho$ (\ref{factor1})
\[
\int ^{l/\sqrt{2}} _{-l\sqrt{2}} T_{11} d x^2 = \frac{1}{4} \int ^{l/2} _{-l/2} (T_{++} +T_{--}) dx^{+}
\]
\begin{equation} 
+ \frac{1}{4}  \int ^{l/2} _{-l/2} (T_{++} +T_{--}) dx^{-} = M_B = \frac{\kappa A}{8\pi}    \label{normalization}
\end{equation}
We have introduced the cutoff parameter $l$ in order to obtain finite results. Eventually we have to take the limit $l \rightarrow \infty$.
from the condition (\ref{normalization}) and from (\ref{factor1}) we obtain for $C$
\begin{equation}
C= \frac{\kappa A}{2\pi l}   \; .      
\end{equation}
We are now ready to define the Virasoro generators by integrating the stress tensor $T_{\pm \pm}$ with some smearing functions $\xi _n$
\begin{equation}
L_n ^{\pm} = \int  dx^{\pm} \xi_n ^{\pm} T_{\pm \pm}  \; ,         \label{viragen1}
\end{equation}
this generators in literature are also called charges. A general charge is composed of two parts $L_n = L_n ^+ + L_n ^- $.
Let us now specify the smearing fields 
\begin{equation}
\xi ^{\pm} _n = \frac{l} {2\pi} \exp \left( \frac{-i n 2\pi x^{\pm}}{l}   \right) \; .           \label{smearing}
\end{equation}
The parameter $l $ is the cutoff already introduced before the factor in front of the exponential is needed to let the fields close 
with respect to the Lie bracket obtaining
\begin{equation}
\left[\xi _n , \xi _m \right] = (n-m) \xi _{n+m}
\end{equation}
The vector fields $\xi$ therefore close the conformal algebra. We want now understand if the variation of the conformal factor induced by those vector fields 
preserve the thermodynamic characteristics of a black hole i.e. it's temperature and entropy.The entropy is proportional to the horizon area of the black hole
and the horizon is located where the lapse function$N$ goes to zero. Therefore if the variation of the lapse $N$ is zero on the horizon also the variation 
of the spatial area of the horizon is zero. Let us therefore compute the Lie derivative
\begin{equation}
\mathcal{L} _{\xi ^{\pm} _n} N(r_0) =0
\end{equation}
This variation is zero on the horizon. Let us now see what happens with the variation of the surface gravity
\begin{equation}
\mathcal{L} _{\xi ^{\pm} _n} \frac{\partial _y N }{\sigma} (r_0) = \kappa \xi ^{\pm} _n \left[  \left( in \frac{2\pi}{l} +1  \right) ^2
- \left( in \frac{2\pi}{l} +1  \right)    \right] \; .
\end{equation}
This expression is not zero for arbitrary $l$, but we have seen that the physically meaningful situation is the one in which $l\rightarrow \infty$
i.e. we push the walls of the box to infinity. In this limit also the variation of the surface gravity tends to zero.
and therefore the fields $\xi$ represent the deformations of the conformal factor which preserve temperature and entropy of the black hole.\\
Let us now notice what happens when we rescale our coordinates by a factor $1/a$ i.e.
\begin{equation}
\tilde{x} = \frac{x}{a}  \;.
\end{equation}
Considering the near horizon solution (\ref{factor1}) we notice that energy of the system does not change under such a rescaling we have in fact
\begin{equation}
\tilde{\partial} _{\pm} \tilde{\rho}(\tilde{x}) = \partial _{\pm} \rho (x) \; ,
\end{equation}
whereas for the square derivative term in the stress tensor we have 
\begin{equation}
\tilde{\partial} _{\pm} \tilde{\partial} _{\pm} \tilde{\rho} (\tilde{x}) = a \partial _{\pm} \partial _{\pm} \rho (x) \; .
\end{equation}
Moreover the vectors $\xi$ transform as
\begin{equation}
\frac{\tilde{\xi} ^{\pm} (\tilde{x})}{a} \tilde{\partial} _{\pm} = \xi^{\pm} \partial _{\pm}  \; .
\end{equation}
In this rescaling the new charges become, discarding for notational simplicity the tildes.
\begin{equation}
L^{\pm} _n = \frac{A \kappa}{2 \pi l} \int _{-l/2 a} ^{l/2 a} (a d x^{\pm}) \frac{l}{2 a \pi} \exp \left( -i \frac{2a\pi}{l} nx^{\pm}  \right)
\left( \partial _{\pm} \rho \partial _{\pm} \rho + \frac{1}{a} \partial _{\pm} \partial _{\pm} \rho  \right)   \; .   \label{newcharge} 
\end{equation}
This rescaling may seem arbitrary. This arbitrariness can be avoided if we impose as condition the matching with Euclidean periodicity
and set so
\begin{equation}
a=\kappa            \; .
\end{equation}
getting so the background metric $ ds^2= \kappa ^2 dx^+ dx^- $. The Euclidean Rindler time $ \tau = -i( x^+ +x^- ) /2 $ has periodicity   
$\frac{2\pi}{\kappa} $ .\\
Notice that the formulas for the Poisson brackets in (\ref{liouvpoisson}) are valid if the constant in front of the stress tensor $C$ is equal to one.
In our case due to the normalization (\ref{normalization}) $c$ is not unity and therefore for dimensional reasons we have to rescale the Poisson bracket
by a factor $-\frac{1}{C}$ namely
\begin{equation}
\{ \dots \} \rightarrow \frac{2\pi l}{A\kappa} \{ \dots \} \; .
\end{equation}
Now we are eventually able to find the Virasoro algebra in fact using (\ref{liouvpoisson}) and (\ref{newcharge}) we can compute
$L_0$ which becomes 
\begin{equation}
L_0 ^+{\pm} =\frac{Al}{16 \pi ^2}    \label{lzero}
\end{equation}
and the Poisson algebra 
\begin{equation}
\left\{ L^{\pm} _n ,  L^{\pm} _m \right\} = i(n-m) L^{\pm} _{m+n} + i \frac{c^{\pm}}{12} n^3 \delta _{m+n}  
\end{equation}
This algebra can be brought in the standard form by a simple shift of the $L_0$ generator.
The central charge in this case is 
\begin{equation}
c^+ = \frac{3 A }{2l}  \; . \label{cplus}
\end{equation}
We have used only one copy of the Virasoro algebra namely the one associated to the future horizon , because the event horizon of the physical
black hole is given by the future horizon.\\
We remember again that eventually we have to make the limit with the cutoff $l$ that tends to infinity. In this case the generator $L_0$ 
diverges and the central charge goes to zero. The final result of zero central charge is therefore in accordance with results in the previous 
chapters. We notice that the product $c^+ L_0$ does not depend on the cutoff $l$. 
We can therefore use the Cardy formula for each finite value of $l$ and then take safely the limit $l\rightarrow \infty$ .
The entropy of the black hole is now given by 
\begin{equation}
2\pi \sqrt{\frac{c^+ L_0 ^+}{6}} = \frac{A}{4}  \; . 
\end{equation}
We obtain therefore the correct Bekenstein-Hawking entropy. Let us notice that the fact that $L_0$ diverges is quite welcome in the use of the
Cardy formula because this formula is an asymptotic formula valid for large $L_0$ values.\\ 
We have studied the conformal factor of a dimensional reduced theory in the presence of a black hole with a
bifurcate Killing horizon.
In this case we have shown that near the horizon, it is described by a Liouville theory, which does not modify the
thermodynamic properties of the black hole. This has a classical central charge and, in fact, we have shown
that it can be used to compute the entropy of the black hole in question. 
If the black hole is extremal it does not posses a bifurcate Killing horizon and in this case the geometry of
the $(r,t)$-plane is not that of a Rindler space but it is an $AdS_2$ space.
In this case our approximation does not hold because 
the exponential term in the equation (\ref{liouville1}) is
positive, and so it is not a Liouville equation.
On the other hand, the temperature of an extremal black hole is zero
We stress the fact that we have  made no assumptions on 
a specific quantum gravity model but used only the classical near-horizon structure of the black hole.
That approach differs from the already studied models because the central role is played by the conformal
field $\rho$ and not by the dilaton $\eta$. 
The effective conformal theory found above describes the micro-canonical theory responsible for the entropy of
the black hole, indeed we have fixed the value of $\int T_{tt}dr$ as its mass $M_B$.
We have found the correct thermodynamic entropy even if the central charge is zero if computed on the global
space, and even if the fundamental mode $L_0^+$ diverges as the cutoff parameter $l$ tends to infinity.
The charges, if computed on the whole space time, correspond to conformal transformations that do not change
the thermodynamic  properties of the black hole. 
On the other hand we have made no assumptions on the boundary condition we have to impose to the conformal
factor. Perhaps some other correction should be searched in that direction.

\section{Generalization to arbitrary dimension}
Up to now we analyzed the dimensional reduction of 4-dimensional spacetime. Today there is quite a big interest also for black
holes in higher dimension. Let us therefore see what happens if we want to generalize  the computation performed in this chapter 
in arbitrary dimension $d$.\\
We start again with the Einstein-Hilbert action this time in $d$ dimensions
\begin{equation}
I_d = \frac{1}{16\pi} \int d^d x \sqrt{-g ^{(d)} } R^{(d)}  \; ,
\end{equation} 
where now $g^{(d)} $ and $R^{(d)}$ are now the d-dimensional metric and scalar curvature.
Again we make the Ansatz of spherical symmetry
\begin{equation}
ds^2 = g_{ij}dx^i dx^j + \Phi ^2 d \Omega _{d-2} ^2 \; ,
\end{equation}
where $g_{ij}$ is the 2-D metric of the $r-t$ plane (for simplicity of notation we will omit the $(2)$ superscripts on the metric and scalar curvature) and
$\Omega_{d-2}$ is the unit $d-2$ sphere.
Now writing the action in function of the 2-D quantities and integrating away the angular degrees of freedom we obtain
\begin{equation}
I = \frac{S_{d-2}}{16\pi} \int \sqrt{g} \left( \Phi ^{d-2} R +(d-3) (d-2)\Phi ^{d-4} (\nabla \Phi )^2 +(d-3)(d-2)\Phi ^{d-4} \right)  \; ,  \label{dreduced}  
\end{equation}
where $S_{d-2}$ is the area of the $d-2$ sphere.
Let us now redefine the dilaton field $\Phi$ as 
\begin{equation}
\Psi = C \Phi ^{d-2} \; \; \; ; \; \; \; C= \frac{S_{d-2}}{2\pi}(\frac{d-3}{d-2}) \; .
\end{equation}
The action (\ref{dreduced}) in terms of $\Psi$ becomes  now
\begin{equation}
I =   \int \sqrt{g} \left( \frac{1}{2} ( \nabla \Psi )^2 +\frac{1}{4}(\frac{d-2}{2(d-3)}) \Psi ^2 R 
+\frac{1}{8} (d-2)^2 C ^{\frac{2}{d-2}}\Psi ^{2(\frac{d-4}{d-2})} \right)
\end{equation}
This is now in a form similar to the action (\ref{reduced1}) in the 4-D case.
For simplicity of notation we will define
\begin{equation}
\left(\frac{d-2}{2(d-3)}\right) \equiv \mathrm{const} \; \; \; ; \; \; \; \frac{1}{8} (d-2)^2 C ^{\frac{2}{d-2}} \equiv \mathrm{konst}
\end{equation}
 As in the 4-D case we redefine the dilaton $\Psi$ and the metric $g$ as
\begin{equation}
\Psi ^2 = \eta \; \; \; ; \; \; \; g= \frac{\beta}{\sqrt{\eta}} \tilde{g}
\end{equation}
The constant $\beta$ is chosen so that we can eliminate the kinetic term of the dilaton in the action. Therefore $\beta$ becomes
\begin{equation}
\beta = \frac{1}{\mathrm{const}}   \; .
\end{equation} 
The action becomes now
\begin{equation}
I = \int \sqrt{\tilde{g}} \left(  \mathrm{const}\frac{\eta}{4}R +\frac{\mathrm{konst}}{\mathrm{const}}\eta ^{\frac{d-6}{2(d-2)}}        \right)   \; ,
\end{equation}
or to put it in a form equal to (\ref{reduced2}) for 4-D 
\begin{equation}
I = \frac{1}{2}\mathrm{const} \int \sqrt{\tilde{g}} \left( \frac{\eta}{2} R + V(\eta )  \right)           \label{dreduced2}
\end{equation}
Where the dilatonic potential is given by
\begin{equation}
V(\eta ) = \frac{2 \, \mathrm{konst}}{(\mathrm{const})^2} \eta ^ {\frac{d-6}{2(d-2)}}   \; .          \label{ddilatonicpotential}
\end{equation} 
Notice that for $d=2$ the potential becomes
\begin{equation}
V(\eta ) = \frac{1}{\sqrt{\eta}}
\end{equation}
and therefore the action becomes exactly the (\ref{reduced2}). We have therefore found the generalization of the dimensionally reduced action 
(\ref{reduced2}) in the case of arbitrary dimension.\\
We can now as before write the metric $\tilde{g}$ in conformally flat form as in (\ref{flatgauge}) and we obtain exactly the action
(\ref{reduced3}) with the potential given by (\ref{ddilatonicpotential}). Let us also notice that passing to light coordinates 
the constraints $T_{\pm \pm}=0$ as given in
 (\ref{constraint3}) do not depend on the potential and therefore we can integrate them as in the 4-D case and also the near 
horizon approximation is independent from the spacetime dimension as it regards only the $r-t$ plane.
Near horizon therefore again we obtain that
\begin{equation}
\partial _{\pm } \eta =0 \rightarrow \eta =\mathrm{const}  \; .
\end{equation} 
Again the only equation that survives is the Liouville equation which now takes the form
\begin{equation}
\partial _+ \partial _- \rho - \frac{\left(\partial _{\eta} V(\eta) \right)_{\eta _0}}{4}e^{-2\rho}=0   \; .
\end{equation}
Therefore the only thing that changes with varying the spacetime dimension is the value of $ \left(\partial _{\eta} V(\eta) \right)_{\eta _0}$,
whose dependence of the dimension is given by (\ref{ddilatonicpotential}).\\
Now in light coordinates the the stress tensor of the Liouville theory does not depend on this constant and therefore also the charges do not.
Therefore neither the value of $L_0$ nor of the central charge change.  We obtain therefore, using the Cardy formula, same result in every dimension i.e.
the correct Bekenstein-Hawking entropy.

\chapter{Free Field}

\section{Integration of the equations of motion}

Let us now return to the equations of motion of the action (\ref{reduced3} without making any near horizon approximation.
We have already integrated the constraints (\ref{constraint3}) obtaining 
\begin{equation}
\partial_{\pm} \eta = \exp \left( -2\rho + C_{\mp}(x_{\mp})  \right) \; ,    \label{arbitrary1}      
\end{equation} 
where the function $C_{\mp} $ is an arbitrary function of $x^\mp$. \\
Let us introduce a function $\phi _{\pm} (x^{\pm} ) $ of $x^{\pm} $ defined as 
\begin{equation}
\partial _{\pm} \phi _{\pm} \equiv \exp{(-C_{\pm})}  \; .   \label{phi}
\end{equation} 
So the eq. (\ref{arbitrary1}) becomes 
\begin{equation}
\partial _{\pm} \eta = e^{-2\rho} \, e^{+C_{\mp}} = e^{-2\rho} \left( e^{-C_{\mp}}  \right) ^{-1} = e^{-2\rho} \left(\partial _{\mp} \phi _{\mp}  \right) ^{-1}
\end{equation}
and therefore we can write 
\begin{equation}
e^{-2\rho} -\partial _{\pm} \eta \partial _{\mp} \phi _{\mp} =0 \; .       \label{integration1}
\end{equation}
Inserting this in the second equation of motion(\ref{motion6}) we obtain
\begin{equation}
\partial _+ \partial _- \eta + \frac{V(\eta)}{2} \partial _{\pm} \eta \partial _{\mp} \phi _{\mp} =0  \; .       \label{integration2}
\end{equation}
Let us now introduce a function of the field $\eta$ $F(\eta )$ defined as 
\begin{equation}
\partial _{\eta}  F(\eta )  \equiv V(\eta)      \label{defeta} 
\end{equation}
Using then the fact that $\phi_{\pm}$ is function only of $x^{\pm}$ we can write (\ref{integration2}) as
\begin{equation}
\partial _+ \partial _- \eta + \partial _{\mp} \left( \frac{ F(\eta ) \partial _{\pm} \phi }{2}  \right) =0 \; . 
\end{equation}
Integrating now with respect to $x^{\mp}$ we obtain
\begin{equation}
\partial _{\pm} \eta + \frac{F(\eta ) \partial _{\pm} \phi _{\pm}}{2} = I_{\pm}    \; ,    \label{integration3}
\end{equation}
where $I_{\pm} $ is a function of $x^{\pm}$ . This function is not completely arbitrary. Let us determinate the form of $I_{\pm}$.
In order to do this consider  (\ref{integration2}), and let us write down explicitly the 2 equations it contains  
\begin{equation}
\begin{array}{c}
\frac{2}{V(\eta )}\partial _+ \partial _- \eta = - \partial _+ \eta \partial _- \phi _- \\
\\
\frac{2}{V(\eta ) } \partial _+ \partial _- \eta = - \partial _- \eta \partial _+ \phi _+ \; .
\end{array}
\end{equation}  
We have then 
\begin{equation}
\partial _+ \phi _+ \partial _- \eta = \partial _- \phi _- \partial _+ \eta  \; .   \label{integration4}       
\end{equation}
Let us now take again (\ref{integration3}) which contains also 2 equations 
\begin{equation}
\begin{array}{c}
\partial _+ \eta + \frac{F(\eta )}{2} \partial _+ \phi _+ = I_+ \\
\\
\partial _- \eta + \frac{F(\eta )}{2}\partial _- \phi _- = I_-
\end{array}
\end{equation}
Let us multiply the first one with $\partial _- \phi _- $ and the second one with with $ \partial _+ \phi _+ $ obtaining
\begin{equation}
\begin{array}{c}
\partial _- \phi _- \partial _+ \eta + \frac{F(\eta )}{2} \partial _+ \phi _+ \partial _- \phi _- =I_+ \partial _- \phi _- \\
\\
\partial _+ \phi _+ \partial _- \eta + \frac{F(\eta )}{2}\partial _- \phi _- \partial _+ \phi _+ =I_- \partial _+ \phi _+
\end{array}
\end{equation}
Using now (\ref{integration4}) we obtain the equation
\begin{equation}
I_+ \partial _- \phi _- = I_- \partial _+ \phi _+  \; .            
\end{equation}
In order to satisfy the last equation the function $\phi$ must have the form 
\begin{equation}
I_{\pm} = C_1 \partial _{\pm} \phi _{\pm}  \; ,        \label{integration5}
\end{equation} 
where $C_1$ is a arbitrary constant.
We have so eventually determined the form of $I_{\pm}$. Using the definition of $F(\eta )$ (\ref{defeta}), we notice that $F(\eta )$ is defined up to an
additive constant. We can therefore redefine $F(\eta )$ as 
\begin{equation}
F_{C_1} = F(\eta ) -2C_1    \; ,      \label{newdefeta}
\end{equation}
where the constant $C_1$ is the one introduced in (\ref{integration5}). Using the redefinition  (\ref{newdefeta}) and (\ref{integration5}) we can eliminate
the function $I_{\pm}$ in (\ref{integration3}) which becomes 
\begin{equation}
\begin{array}{c}
\partial _+ \eta = - \frac{F_{C_1}}{2} \partial _+ \phi _+\\
\\
\partial _- \eta = - \frac{F_{C_1}}{2} \partial _- \phi _-
\end{array}            \label{eliminatei}
\end{equation}
which can be integrated 
\begin{equation}
\begin{array}{c}
2\int \frac{\partial _+ \eta }{F_{C_1} (\eta)}dx^+ = -\phi _+ \\
\\
2\int \frac{ \partial _- \eta}{F_{C_1} (\eta )} dx^- =- \phi _- \; .
\end{array}                              \label{integration6}
\end{equation}
Let us now define the field $\phi$ as
\begin{equation}
\phi \equiv \phi _+ + \phi _-   \; .           \label{defphi}
\end{equation} 
Using the fact that
\begin{equation}
d\eta (x^+ , x^- ) = \partial _+ \eta dx^+ + \partial _- \eta dx^-
\end{equation}
equations (\ref{integration6}) can be written as
\begin{equation}
-\phi = 2\int \frac{d\eta}{F_{C_1}(\eta )} +C_2  \equiv 2 G_{C_1 , C_2}  \; ,       \label{integration7} 
\end{equation}
where $C_2$ is an integration constant. The function $\phi$ as defined in   (\ref{defphi}) satisfies obviously the free field equation
\begin{equation}
\Box \phi =0      \label{freefield}  \; .  
\end{equation}
We have expressed the dilaton field $\eta$ in function of a free field
We can also express the Liouville field $\rho $ in function of $\phi$. Using in fact (\ref{integration1}) and (\ref{eliminatei}) we obtain
\begin{equation}
e^{-2\rho} = \partial _{\pm}\phi \partial _{\mp} \eta = -\frac{F_{C_1}(\eta )}{2} \partial _{\pm} \phi \partial _{\mp} \phi
\end{equation} 
Using (\ref{integration4}) we see that we need to keep only one of the two equations e.g.
\begin{equation}
e^{-2\rho} =- \frac{F_{C_1}(\eta ) }{2}\partial _+\phi \partial _-\phi   \; .  \label{integration8}
\end{equation} 
We started from the equations of motion and the constraints derived from the action (\ref{reduced3}) without any near horizon approximation,
which involved 2 fields namely $\eta$ and $\rho$. We have shown that this two fields can be expressed by means of one free field. 
Summarizing we obtained
\begin{equation}
\begin{array}{c}
\Box \phi =0  \\
\\
-\phi = 2 G_{C_1 , C_2} (\eta)\\
\\
e^{-2\rho} = - \frac{F_{C_1}(\eta ) }{2} \partial _+ \phi \partial _- \phi
\end{array}                 \label{freefield2}
\end{equation}

\section{Constant potential case}
In this chapter we have made up to now no near horizon approximation. We have used only the Ansatz of spherical symmetry.
In the previous chapter we have seen that one consequence of the near horizon approximation was, that in this case the dilaton is almost constant near the horizon
and therefore also the dilatonic potential can be considered constant i.e. $V(\eta) = V(\eta _0 )$.\\
We want now to see in the constant potential case what form the free field must have in order to obtain from (\ref{freefield2}) for the Liouville field
$\rho$ the Rindler form (\ref{factor1}), which describes the near horizon geometry. 
Taking
\begin{equation}
V(\eta ) = \lambda  = \mathrm{const} \;.
\end{equation}
Therefore from (\ref{defeta}) we get
\begin{equation}
F(\eta ) = \lambda \eta \; .
\end{equation}
Using (\ref{freefield2}) 
\begin{equation}
G(\eta ) = \frac{1}{\lambda} \log (\lambda \eta) \; .
\end{equation}
Let us now take a free field solution for $\phi$. A possible choice is e.g.
\begin{equation}
\phi = ax^+ - bx^- \;,
\end{equation}
with constant $a$ and $b$. Using this choice from the second equation of (\ref{freefield2}) we get
\begin{equation}
\frac{2}{\lambda} \log (\lambda \eta ) = - \phi = -(ax^+ - bx^- )
\end{equation}
and so we get for $\eta$
\begin{equation}
\eta = \frac{1}{\lambda} \exp \left( - \frac{\lambda}{2} [ax^+ - bx^- ] \right)  \; .
\end{equation}
Taking now the third equation of (\ref{freefield2}) we obtain
\begin{equation}
\exp (-2\rho ) = \frac{ \lambda \, \eta}{2} \, a b = \frac{ab}{2} \exp \left( \frac{\lambda}{2}[ax^+ - bx^- ] \right) \; ,  
\end{equation}
Taking now for example $a=b=\sqrt{2}$ we have
\begin{equation}
\exp (-2\rho ) = \exp \left( \frac{\lambda \sqrt{2}}{2}[x^+ - x^- ] \right) 
\end{equation}
Comparing the exponents in the last equation we see that, with this choice of the free field $\phi$, the Liouville field $\rho$ has the
Rindler form (\ref{factor1}). Moreover the fields $\rho$ and $\phi$ are proportional and being linear in the coordinates they are related 
by a scale transformation. Being a free scalar field theory in two dimensions conformally invariant the two fields $\rho $ and $\phi$ are to be considered
physically equivalent. Therefore we argue that the near horizon physics of a black hole can be described completely by a free field.\\
In order to have a complete description of the black hole microstates we must eventually quantize theory describing them. Having a free field 
describing the microstates  the discussion of some quantum issues of black hole will be possible. The fact that $2-D$ quantum gravity may be described 
by a free field was noticed also in a different context \cite{polyakov2}.

\section{Coupling to the gravitational field} 
We have seen up to now that the equations of motion of the dimensional reduced, action in the case of spherical symmetry can be integrated in terms
of a free field $\phi$. In the near horizon approximation the form of this free field, that gives the near horizon solution
(\ref{factor1}) for the conformal factor $\rho$, is up to a rescaling of the coordinates equal to the near horizon solution for $\rho$.
Due to conformal invariance the two fields are therefore equivalent.\\
Let us therefore have a look at the free field theory. The action for the free field  $\phi$, which propagates in flat spacetime is of the form
\begin{equation}
I_{free} = C \int d^2x \gamma ^{\mu \nu} \partial _{\mu} \phi \partial _{\nu} \phi  \label{freescalar2}
\end{equation}
The constant $C$ is arbitrary because the integration of the eq. of motion in (\ref{freefield2}) require only that $\Box \phi =0$. 
We have already mentioned that in two dimensions and only there the scalar free field action is conformally invariant. 
Using the action (\ref{freescalar2}) we are therefore supposing that our field $\phi$ transforms like a scalar.
The stress tensor derived form this action is 
\begin{equation}
\frac{1}{C} T_{\mu \nu} = \partial _{\mu} \phi \partial_{\nu} \phi - \frac{1}{2} \gamma _{\mu \nu} \partial ^{\alpha} \phi  \partial _{\alpha} \phi  \; ,
\end{equation} 
which has zero trace.\\
Notice now that the the background geometry is flat. If we want to couple the free field to a non-flat background we can use the action
\begin{equation}
I_c =C \int d^2x g^{\mu \nu} \nabla _{\mu} \phi \nabla _{\nu} \phi + \alpha R[g]\phi   \label{coupling2}   \; .
\end{equation}
The background metric $g$ is here taken arbitrary. Taking now as choice for the background metric the flat metric $g= \gamma$,
the scalar curvature of course is zero. In the flat background therefore the two actions (\ref{freescalar2}) and (\ref{coupling2}) 
give the same equation of motion.  The difference in the two actions is in the stress tensor. The stress tensor of (\ref{coupling2})
must in fact be computed before choosing the form of $g$. Using the formula for the variation of the $R\phi$ term in the appendix 
we obtain the improved stress energy tensor
\begin{equation}
\frac{1}{C} T_{\mu \nu} = \partial _{\mu} \phi \partial_{\nu} \phi - \frac{1}{2} \gamma _{\mu \nu} \partial ^{\alpha} \phi  \partial _{\alpha} \phi 
+ \alpha \left( \gamma _{\mu \nu} \Box \phi - \partial _{\mu} \partial _{\nu}  \right)
                                                                                      \label{improved2}
\end{equation}
The term proportional to $\alpha$ comes from the $R\phi$ coupling and remains also when we take the flat background.
Let us check the trace of the improved stress energy tensor with general background $g$ 
\begin{equation}
T^{\mu} _{\mu} = \nabla _{\mu} \phi \nabla ^{\mu} \phi -\frac{2}{2} \nabla _{\mu}\phi \nabla ^{\nu}\phi +2 \alpha \Box \phi - \alpha\Box \phi= \alpha \Box \phi 
\end{equation} 
The equation of motion for $\phi$ from the action (\ref{coupling2}) in arbitrary background is
\begin{equation}
\Box \phi =\alpha R[g]
\end{equation}
inserting it in the trace formula we have  
\begin{equation}
T^{\mu} _{\mu} = \alpha ^2 R[g]  \; .
\end{equation}
There is therefore formally a trace anomaly in the improved stress energy tensor proportional to $\alpha ^2$. Taking $g=\gamma$ the trace becomes zero.
Therefore in the flat background  for the stress energy tensor of action (\ref{freescalar2}) is identically zero, 
whereas the trace of the improved stress energy tensor is zero on shell.\\
Let us now notice that the action (\ref{coupling2}) is not Weyl invariant if $\phi$ transforms like a scalar. Using the calculations in (\ref{appendixweyl})
we see that the action (\ref{coupling2}) is Weyl invariant if the field $\phi$ transforms not like a scalar but like the Liouville field i.e.
under a Weyl transformation parametrized as 
\begin{equation}
g_{\mu \nu} \rightarrow e^{\omega} g_{\mu \nu}    \label{bla1}
\end{equation}
the field $\phi$ shifts as 
\begin{equation}
\phi \rightarrow \phi + \beta \omega         \label{bla2}
\end{equation}
We want to determinate the constant $\beta $ for which we have Weyl invariance of the action (\ref{coupling2}). Using the Ansatz (\ref{bla1}) and
(\ref{bla2}) and using the  calculations of (\ref{appendixweyl}) the action (\ref{coupling2}) becomes
\[
I' _c = C\int \sqrt{g}\left[ \left(\nabla \phi \right)^2 + \alpha \phi R[g] - 2\beta \nabla \phi \nabla \omega - \alpha \phi \nabla ^2 \omega \right.
\]
\begin{equation}
 \left. +\left( \nabla \beta \omega  \right)^2 - \alpha \beta \omega R[g] + \alpha \beta \omega \nabla ^2 \omega  \right]    \; .
\end{equation} 
The last three terms do not contain the field $\phi$ and therefore don't affect the equation of motion and can therefore be discarded.
The first two terms are the originary action. The only dangerous terms are therefore
\begin{equation}
-2\beta \nabla \phi \nabla \omega\stackrel{partial \, int.}{\rightarrow} +2\beta \phi \nabla ^2 \omega 
\end{equation}
\begin{equation}
- \alpha \phi \nabla ^2 \omega   \; .
\end{equation}
In order to have Weyl invariance the two terms must cancel and therefore we obtain
\begin{equation}
\phi \nabla ^2 \omega \left( 2\beta -\alpha  \right)=0 \rightarrow \beta = \frac{\alpha}{2}
\end{equation}
and therefore the transformation law for $\phi$ is 
\begin{equation}
\phi \rightarrow \phi - \frac{\alpha}{2} \omega  \; .      \label{newshift}
\end{equation}
Therefore the action (\ref{coupling2}) is Weyl invariant for every value of the coupling constant $\lambda$ if the field transforms 
according to (\ref{newshift}).\\
This is different from the Liouville case where in the action we have also the exponential term
$e^{\gamma\phi}$. In this case in order to have Weyl invariance the coupling constant $Q$ of the $r\phi$ coupling in (\ref{liouvaction})
must take the value $Q= \frac{2}{\gamma}$.  \\
We have therefore seen up to now that the two actions (\ref{freescalar2}) and (\ref{coupling2}) in the flat gauge give the same equation of motion
but different stress tensors and different transformation laws for the field $\phi$.
The question is now which of the two actions for $\phi$ is the correct one for the description of black hole dynamics.
We have already said that in order to have a complete theory of the black hole microstates we must eventually quantize our theory.
Our field $\phi$ moves in a flat background, but in the  quantum theory there will be also fluctuations of the background metric and 
therefore it seems natural to include a $\phi R$ coupling.\\
On the other side we have seen that metric of the $r-t$ plane of the black hole and therefore all the relevant part of it's geometry
is described by means of the conformal factor field  $\rho$ which behaves not like a scalar. In the last section we have seen that
in the near horizon limit the field $\phi$ is physically equivalent to $\rho$ differing from it only by a rescaling. We can therefore conclude that  
also $\phi$ should not transform like a scalar but like (\ref{newshift}). It seems therefore reasonable to choose the action
(\ref{coupling2}) for the field $\phi$.\\
Let us now write the improved stress tensor in light coordinates we obtain
\begin{equation}
T_{\pm \pm} = C\left( (\partial _{\pm} \phi )^2 + \alpha \partial ^2 _{\pm} \phi  \right)   \; .
\end{equation}    
This is identical to the Liouville stress tensor. We know therefore that in the Poisson algebra we has then a central extension  in  the form of 
(\ref{liouvpoisson}) 
\begin{equation}
\left\{ Q_f , Q_g    \right\} = Q_{[f,g]} + \frac{\alpha ^2}{4 } \Delta (f,g) 
\end{equation}
and therefore with a central charge proportional to $\alpha ^2$ which is not a surprise because of the transformation law (\ref{newshift}) for $\phi$.
Using the action (\ref{freescalar2}) the stress tensor in light coordinates has the form
\begin{equation}
T_{\pm \pm} =C (\partial _{\pm} \phi )^2
\end{equation}
and therefore there is no central charge. In this case in fact the field $\phi$ is a real scalar field.\\
The fact that the coupling $\alpha$ is completely arbitrary can be used to set the central charge in such a way to cancel the 
ghost contribution, which will be done in the next section.

\section{Cancellation of the ghosts contributions}
If we quantize our theory with described by the action (\ref{coupling2})  the improved energy stress tensor becomes an operator.
the central charge can then be computed performing an operator product expansion of $T$. Using complex coordinates it has the form
\begin{equation}
T(z)T(w) = \frac{c}{2} \frac{1}{(z-w)^4} +\frac{2}{(z-w)^2} T(w) + \frac{1}{z-w}\partial _w T(w)
\end{equation} 
The part proportional to $c$ gives the anomaly, i.e. the non-tensor piece in the transformation of $T$. The  central charge takes the value
 \cite{polchinski}, \cite{ginsparg}
\begin{equation}
c = 1 + 3\alpha ^2         \; .   \label{quantumcentralcharge}
\end{equation}
The term proportional to $\alpha ^2 $ is the classical contribution to the central charge seen before. The other term is a quantum correction to 
the central charge.\\
Let us notice that the field $\phi$ propagates in a flat background because we have used the conformal flat gauge.
The quantization can be done using the path integral formalism defining the quantity
\begin{equation}
Z =\int d \phi d g \exp (-I_e)          \label{pathintegral}
\end{equation}
where the action $I_e$ is the Euclidean version of (\ref{coupling2}) and the path integral runs over all possible Euclidean metrics $g$. The fact that
we pass to the Euclidean metric is needed for the convergence of the path integral.
The definition (\ref{pathintegral} as written so is not completely correct because configurations related by a $Diff \times Weyl$ symmetry are 
physically equivalent. The sum over configurations is therefore meaned as modulo    $Diff \times Weyl$ group.
This can be done by fixing a gauge of the metric e.g. the flat metric as we have done in the classical case at the price of introducing ghost fields.
The path integral can be written as \cite{polchinski}
\begin{equation}
Z[\hat{g}] = \int d\phi d b dc \exp (-I_e - I_{gh})   \; ,               \label{pathghost}
\end{equation}
where the action $I_{gh}$ is the action of the ghost fields and $\hat{g}$ means the gauge fixed metric.
Having gauge fixed the metric now it is crucial to check if the path integral is independent of the gauge choice i.e.
\begin{equation}
Z[\hat{g}] = Z[\tilde{g}]   \; ,
\end{equation}
where the metrics $\hat{g}$ and $\tilde{g}$ are related by a  Weyl transformation. Let us therefore make a Weyl transformation with
$\delta g = \omega g$. We obtain as variation of $Z$
\begin{equation}
\delta Z[g] = \int d^2 x \sqrt{g} T^{ab}\delta g_{ab} \int d\phi db dc \exp ( -I_e -I_{gh})  
\end{equation}
Using now the Weyl form form $\delta g_{ab}$ we obtain 
\begin{equation}
 \delta Z[g] = \int d^2 x \sqrt{g} \omega T^a _a  \int d\phi db dc \exp ( -I_e -I_{gh})    \; . 
\end{equation}
This expression must be zero if we want gauge independence.
We conclude that in order to have a background metric independent path integral the stress tensor must have zero trace, where in the quantum case 
the stress tensor is of course an operator. The last equation is therefore the operator generalization of the classical Weyl invariance condition
\begin{equation}
T^a _a =0 
\end{equation}
It is a standard result \cite{polchinski} that a possible trace anomaly must be of the form
\begin{equation}
t^a _a = \frac{c}{12} R  \; . 
\end{equation} 
Therefore if the total central charge is not zero there will be a trace anomaly and therefore different gauge choices would be inequivalent.
Notice that in the classical case the trace of the action was zero on shell also if we had a nonzero central charge because the used the flat 
background.\\
We have now three contributions to the central charge: a classical proportional to $\alpha ^2$ a quantum correction (see eq. (\ref{quantumcentralcharge}))   
and a contribution from the ghost fields introduced in (\ref{pathghost}
\begin{equation}
c_{gh} = -26      \; .
\end{equation}
As said before  in the quantized theory that all the contributions to the central charge must  cancel eventually obtaining 
\begin{equation}
c_{class} + c_{quant} + c_{gh} = 0
\end{equation}
We can therefore determinate the value of the parameter $\alpha $ in order to have a consistent theory which becomes 
\begin{equation}
\alpha ^2 = \frac{25}{3}
\end{equation}
In this way the path integral based on the action (\ref{coupling2}) will be Weyl invariant. This action seem therefore the suitable starting point
for a quantum theory describing the black hole microstates.

\chapter*{Conclusions}
In this thesis we have studied the possibility to count the microstates responsible for the entropy of the black hole, without
using details of a specific model of quantum theory of gravity (e.g. superstrings, quantum geometry etc.).\\ 
This was done by using a classical symmetry principle, namely two dimensional conformal symmetry. This symmetry in fact is strong 
enough to fix the asymptotic behavior of the density of the microscopic degrees of freedom by means of the Cardy formula.
The Cardy formula uses only the $L_0$ eigenvalue and the central charge $c$ of the generator algebra.\\
The Cardy formula usually arises in a quantized two dimensional CFT, where the central extension of the conformal algebra 
arises as quantum anomaly. We have seen that also at classical level, in the Poisson algebra of the canonical generators, 
a nontrivial central extension of the conformal algebra may arise (chapter 2).\\
We have studied two possible origins of such nontrivial classical central extensions. 
The first possible origin of a classical central extension described in chapter 3,
was due to the boundary terms that one has to add to the canonical generators of diffeomorphisms in order to make them 
differentiable and so to be able to define the Poisson brackets. In fact the boundary terms are defined up to arbitrary 
functions that do not depend on the canonical variables. This means that the generator, resulting from the Poisson bracket 
may not match this arbitrary function. 
The form of the  boundary terms depends on the exact boundary condition that one imposes. In the case of a static black hole with standard foliation
all the spacelike  hypersurfaces intersect in the bifurcation, which becomes so an inner boundary.   \\
We have studied two possible boundary conditions on the bifurcation,  namely fixed bolt metric and fixed surface gravity on the bolt.
In the case of fixed metric on the bolt the action acquires a boundary term associated to it.
Performing the Legendre transformation one sees that the boundary term associated to the bifurcation is canceled and therefore 
the hamiltonian and the canonical generators do not have a boundary term associated to the bolt.
This means that the generator algebra becomes the constraints algebra which does not admit central charges.\\
In the case of fixed surface gravity we have seen that the action has no boundary term regarding the bolt, but in this case the 
hamiltonian and the canonical generators acquire a boundary term. We noticed that the diffeomorphisms must also preserve the existence of the bolt
and not only the surface gravity.
Having a boundary term we can compute on shell the Poisson brackets. One discovers that there is no nontrivial central term in the
generator algebra
So unfortunately also in the case of fixed surface gravity there is no central charge and so the Cardy formula cannot be used.\\
This does not mean that it is impossible to find a central charge. In fact in chapter 4 we have seen that there can be also a second 
origin for a classical central charge, namely if we have a field that does not transform like a scalar, like the Liouville field.
We performed therefore the dimensional reduction of the Einstein-Hilbert under the Ansatz of spherical symmetry. We also wrote
the two dimensional metric of the $r-t$ plane in conformal flat form obtaining an effective theory with two fields, namely the dilaton and the conformal factor.
Using the near horizon approximation the dilaton freezes out and the equation of motion for the conformal factor becomes the Liouville equation,
which is a conformal field theory.
The conformal factor being a piece of a metric has in fact the same anomalous transformation property as the Liouville field.        
Therefore we obtain a central charge for this theory. Using the Cardy formula we obtained exactly the Bekenstein-Hawking entropy.
We can conclude therefore that the we can count the microscopic degrees of freedom of the black hole by means of the Cardy formula
because the effective near horizon field theory is described by a Liouville field which has an anomalous transformation law under conformal 
transformations.\\
In chapter 5 we have then seen that the effective 2-D theory describing the black hole is equivalent to a free field theory.
We have seen that this free field propagating in a flat spacetime can be coupled to the scalar curvature without modifying the 
equations of motion, being the scalar curvature zero. What changes is the stress energy tensor which gets a nonzero contribution from the 
scalar curvature also if we use then the flat gauge. The action with the coupling to the scalar curvature is Weyl invariant only if 
the field transforms not like a scalar but like Liouville field. Having seen, that the Liouville field is near horizon is equivalent to the
free field, it is physically meaningful, that also the free field transforms like the Liouville field.\\
Having this transformation property again the Poisson algebra of this theory acquires a central charge. Now in order to have a complete
theory of the black hole microstates we must eventually quantize the theory. We have noticed that the coupling constant for the $\phi R$
coupling is completely arbitrary. This arbitrariness can be used to cancel the ghosts contribution coming from the path integral
so that the total central eventually is zero. In this way the path integral is independent from the gauge choice and so we have a consistent theory.

\addcontentsline{toc}{chapter}{Conclusions}

\appendix

\chapter{Weyl invariance of the Liouville action}              \label{appendixweyl}

Let us consider the Liouville action (\ref{liouvaction})
\begin{equation}
I_{Liouv} = \frac{1}{8\pi} \int d^2 x \sqrt{\hat{g}}\left[ (\hat{\bigtriangledown} \phi)^2 + QR[\hat{g}]\phi +\frac{\mu}{\gamma ^2}
e^{\gamma \phi} \right]   \; .
\end{equation}
And let us see if it is invariant under Weyl transformations (\ref{weyl})
which here we parametrize as
\begin{equation}
 \hat{g} \rightarrow e^{\omega} \hat{g}
\end{equation}
 combined with  the shift of the $\phi$ (\ref{shift}). We obtain
\[
I' = \frac{1}{8\pi} \int d^2 x \,  \sqrt{\hat{g}} \, e^{\omega} \left[ \left( \hat{ \nabla} \left[ \phi - \frac{\omega}{\gamma} \right] \right) ^2 \right.
\]
\begin{equation}
+ \left. \left.  Q\left( \phi - \frac{\omega}{\gamma} \right) e^{-\omega} \left( R [ \hat{g} - \hat{ \nabla} ^2 (\omega)  \right) 
 +\frac{ \mu}{ \gamma ^2} e^{ \gamma \phi -\omega} \right.  \right]  \; ,     \label{conto1}     
\end{equation}
where in the last line we have used the transformation property of the 2-D scalar curvature under conformal transformations (\ref{curvatureconform}).
Now in order to continue the computation let us notice that for function $f$ we have
\begin{equation}
\left( \hat{ \nabla} f  \right) ^2 = \hat{g} ^{\mu \nu} \hat{ \nabla} _{\mu} f \hat{ \nabla} _{\nu} f
\end{equation}
and therefore under the Weyl transformation $ \hat{g}^{\mu \nu} \rightarrow e^{-\omega} \hat{g}^{\mu \nu}$ we obtain 
\begin{equation}
\left(\hat{\nabla} f  \right) ^2 \rightarrow e^{-\omega} \left(\hat{ \nabla} \phi  \right)
\end{equation}
inserting this in eq. (\ref{conto1}) we get

\[
 I' = \frac{1}{8\pi} \int d^x \, \sqrt{\hat{g}} \left[ \left( \hat{\nabla} \phi  \right) ^2  - 2 \hat{ \nabla} \phi \hat{\nabla} \left( 
\frac{\omega}{\gamma} \right) + \underbrace{\hat{\nabla} \left( \frac{\omega}{\gamma} \right) ^2} + Q\phi R[\hat{g}]  \right.
\]
\begin{equation}
\left.  - Q\phi \hat{\nabla} ^2 (\omega) - \underbrace{\frac{\omega}{\gamma}R[\hat{g}]Q   } + \underbrace{\frac{\omega}{\gamma}Q \hat{\nabla} ^2(\omega)  }
+ \frac{\mu}{\gamma ^2} e^{\gamma \phi}     \right]   \; .
\end{equation}
The underbraced terms do not depend on the Liouville field $\phi$ and so don't change the equation of motion. We can therefore 
discard them.\\
The action $I'$ is to equal to the Liouville action $I_{Liouv}$ only if the two terms depending on the field $\rho$ in $I'$  cancel i.e. 
\begin{equation}
-2 \hat{\nabla} \phi \hat{\nabla} \left( \frac{\omega}{\gamma}\right) - Q \phi \hat{\nabla} ^2 (\omega) = 0 \;.    \label{conto2} 
\end{equation}
Integrating now by parts the first term of the last equation and discarding the total divergence eq. (\ref{conto2}) becomes
\begin{equation}
2\phi \hat{\nabla} ^2 \left( \frac{\omega}{\gamma} \right) - Q \phi \hat{\nabla} ^2 (\omega ) = 0    \; .
\end{equation}    
We can therefore conclude that the Liouville action is Weyl invariant only if 
\begin{equation}
Q = \frac{2}{\gamma}     \; .
\end{equation}

\chapter{ Variation of $R$ coupled to Liouville field in $D=2$}

Let us take for a two dimensional spacetime the action of the form
\begin{equation}
S= \int \sqrt{g}\, \phi R = \int \sqrt{g}\, g^{ab} R_{ab} \phi     \label{coupling}
\end{equation}
We want to compute the variation of (\ref{coupling}) with respect to the metric
\begin{equation}
\delta S = \int \left( \left( \delta  \sqrt{g} \right)g^{ab} R_{ab}  \right)\phi + \int \left( \sqrt{g}\left( \delta g^{ab}  \right) R_{ab}  \right) \phi
+\int \left( \sqrt{g} g^{ab} \delta R_{ab}  \right) \phi              \label{varcoupling}
\end{equation}
Now the variation of $\sqrt{g}$ is \cite{wald}
\[
\delta \sqrt{g} =  -\frac{1}{2}\sqrt{g} \, g_{ab} \, \delta g^{ab}
\]
Therefore we can write (\ref{varcoupling}) as
\begin{equation}
\delta S = \int \phi \left[ \sqrt{g}\left( R_{ab} -\frac{1}{2}g_{ab} \, R  \right)  \right] \delta g^{ab} +\int \phi \, \sqrt{g} g^{ab} \delta R_{ab}
\end{equation}
The first term in the last equation is proportional to the Einstein tensor. It is well known that in $D=2$ this is identically zero \cite{jackiw2} \cite{collas}.
therefore survives only
\begin{equation}
\delta S = \int \phi \, \sqrt{g} \, g^{ab} \delta R_{ab}        \label{varcoupling2}
\end{equation}
As usual the variation of the scalar curvature is a total divergence \cite{wald}
\begin{equation}
g^{ab} \, \delta R_{ab} =  \nabla ^a v_a             \label{rvariation}
\end{equation}
with
\begin{equation}
v_a = \nabla ^b \delta g_{ab} - g^{cd} \nabla _a \delta g_{cd}
\end{equation}
Now in our case the scalar curvature is coupled to the Liouville field and therefore eq. (\ref{varcoupling2})
isn't  simply a boundary term. 
Let us compute therefore
\begin{equation}
=\phi \, g^{ab} \delta \, R_{ab} = \phi \, \nabla ^a v_a = \phi \, \nabla ^a \left[ \nabla^b \delta g_{ab} - g^{cd} \,  \nabla _a \delta g_{cd}  \right]
\end{equation} 
Using the fact that the covariant derivative of the metric is zero we obtain
\begin{equation}
\phi \left[ \nabla ^a \nabla ^b \delta g_{ab} - g^{cd}\nabla ^a \nabla _a \delta g_{cd}  \right]
\end{equation}
integrating by  parts gives 
\[
=  \left( \nabla ^a \left[ \phi \nabla ^b \delta g_{ab} \right] - \nabla ^a \phi \nabla ^b \delta g_{ab}
- g^{cd} \nabla ^a \left[ \phi \nabla _a \delta _{cd}  \right] + g^{cd} \nabla ^a \phi \nabla _a \delta g_{cd} \right)
\]
Integrating again by parts gives 
\[
=  \nabla^a \left[ \phi \nabla ^b \delta g_{ab} \right] - \nabla ^a \nabla ^b \left[ \phi \delta g_{ab}  \right]
+ \left[ \nabla ^a \nabla ^b \phi \right] \delta g_{ab} - g^{cd} \nabla ^a \left[ \phi \nabla _a \delta g_{cd}  \right]                              
\]
\[
+ g^{cd} \nabla ^a \nabla _a \left[ \phi \delta g_{cd}  \right] - g^{cd} \,  \Box \phi \delta g_{cd} 
\]
obtaining eventually
\begin{equation} 
= \left( \nabla ^a \nabla ^b \phi - g^{ab} \Box \phi \right) \delta g_{ab} = \delta \int \sqrt{g} \, \phi \, R = \delta S    \label{varcoupling3}
\end{equation}

\chapter{ADM decomposition of the metric}
The hamiltonian formalism describes the evolution of space quantities in time. Therefore for the hamiltonian description of general relativity we must
introduce a foliation of spacetime, i.e. a family of spacelike hypersurfaces $\Sigma _t$ for which for every point of spacetime $x \exists ! \overline{t} $ such
that $x \in \Sigma _{\overline{t}}$. The standard case is given by constant time hypersurfaces, which will be used here.  
On every hypersurface $\Sigma$ the 4-metric $g_{\mu \nu}$ induces a spatial metric \cite{hawking&ellis}  
\begin{equation}
h_{ab} = g_{ab} + u_a u_b \;,
\end{equation}
where the  vector  $u$ is the normal to $\Sigma$. The metric $h_{ab}$ is also called first fundamental form.   
Given $h_{ab} $ alone it is not possible to reconstruct the spacetime metric. We must also now the embedding of the spacelike hypersurfaces in spacetime.
Let us take two hypersurfaces $\Sigma _{t_1}$ and $\Sigma _{t_2}$ with $t_2 = t_1 +dt$. The proper time $\tau$ elapsed going from $\Sigma _{t_1}$ to 
$\Sigma _{t_2}$ along the hypersurface normal $u$ will be proportional to  $dx^0$ 
\begin{equation}
d\tau = N (x^0 , x^i) dx^0  \; .       \label{lapsefunction}
\end{equation}
The function $N$ is called lapse function. In general the time flow vector is not orthogonal to $\Sigma _t$. Therefore moving from a point $p$ with
coordinates $(x^i)$ on  $\Sigma _{t_1}$  along the hypersurface normal $u$ we will arrive on $\Sigma _{t_2}$ on a point $p'$ with coordinates  
$(x ^{i}) ' \neq x^i $. There is a shift proportional to $dx^0$
\begin{equation}
\Delta x^i = N^i (x^0 , x^i) dx^0 \; ,         \label{shiftvector}
\end{equation}
where the vector $N^i$ is called shift vector.\\
Making now a generic displacement from one hypersurface to the other i.e. from $(x^0 , x^i )$ to $x^0 +dx^o , x^i +dx^i$ we can decompose it in a part normal
and tangent to $\Sigma$. The total tangent displacement shift $d ^{tot} x^i$ is composed of a piece due to the spacetime embedding (\ref{shiftvector})    
and of a piece due to the shift in the chosen direction.
\begin{equation}
d^{tot} x^i = \Delta x^i + dx^i = N^idx^0  +dx^i         \label{totalshift} 
\end{equation}
Using now (\ref{lapsefunction}) the total spacetime interval is given by
\begin{equation}
ds^2 = h_{ik} \left( N^i dx^0 +dx^i \right) \left( N^k dx^0 + dx^k   \right)  -\left( Ndx^0  \right) ^2  \; .  \label{decompose1}
\end{equation}
We have also the relation 
\begin{equation}
N^i = h^{ij} N_j       \label{index}
\end{equation}
Using (\ref{decompose1}) and (\ref{index}) we obtain the decomposition of the metric
\begin{equation}
ds^2 = h_{ik} dx^i dx^k + 2N_i dx^i dx^0 + \left( N^i N_i - N^2  \right)(dx^0)^2   
\end{equation}
or in matrix form
\begin{equation}
 g_{\mu \nu}=  \left(  \begin{array}{cc}
(N^i N_i -N^2 ) & N_k \\
N_i             & h_{ik} 
\end{array}    \right)
\end{equation} 
and the inverse metric is
\begin{equation}
g^{\mu \nu} = \left( \begin{array}{cc}  
-\frac{1}{N^2}  & \frac{N^m }{N^2}\\
\frac{N^k}{N^2} & \left[ h^{km} - \left(\frac{ N^k N^m}{N^2}   \right)  \right]
\end{array}   \right)
\end{equation}
The embedding of the hypersurfaces is described by $N$ and $N^i$.\\
We want to describe the embedding by means of a tensor defined on $\Sigma$. To do this notice that the embedding of $\Sigma$ can be described by the
variation of the normal vector $u$ from point to point on $\Sigma$. Going now from a point $x$ on $\Sigma $ to a point $x +\delta x $ for the 
variation of $u$ we make the Ansatz
\begin{equation}
\delta u ^p = K ^p _q \delta x^q       \label{defextrinsic} 
\end{equation}  
The components of $u$ are
\begin{equation}
u_0 = N \; \; \; ; \; \; \; u_i =0
\end{equation}
or 
\begin{equation}
u^0 = \frac{1}{N} \; \; \; ; \; \; \; u^i = \frac{N^i}{N}  \; .
\end{equation}
Using now the 4- geometry and parallel transporting  $u$ from $x$ to $x+ \delta x$ on $\Sigma$ 
\[
\delta u_i = u_{i;k}\delta x^k
\]
\begin{equation}
= \left( \partial _k u_i - \Gamma ^{\sigma} _{ik} u_{\sigma}    \right)\delta x^k = -N\Gamma ^{0} _{ik} \delta x^k  \; .        \label{extrinsic1}
\end{equation}
The semicolon and $\Gamma$ denote the covariant derivative and Christoffel symbols with respect to the 4-metric.
Comparing (\ref{defextrinsic}) and (\ref{extrinsic1}) we obtain
\begin{equation}
K_{ik} = n_{i;k} = -N\Gamma ^{0} _{ik} \; ,
\end{equation}
which can be written as 
\begin{equation}
K_{ik} =\frac{1}{2N} \left( N _{i/k} + N_{k/i} - \partial _t h_{ik}  \right)    \; .
\end{equation}
The symbol $/$ denotes spatial covariant derivative. The spatial tensor $K_{ij} $ is called extrinsic curvature which can also be written as
\begin{equation}
K_{ij} = \frac{1}{2} \mathcal{L}_u h_{ij}
\end{equation}

\chapter{Hamiltonian constraints}

Let us consider the the bulk term of the gravitational hamiltonian 
\begin{equation}
 H = \int d^3x N \mathcal{H} +\int d^3 x N^i \mathcal{H} _i 
\end{equation}
The functions $\mathcal{H} $ and $\mathcal{H} _i $ are given by 

\begin{equation}
\mathcal{H} = \frac{16\pi}{\sqrt{h}}\left[ P_{ab}P^{ab} - \frac{1}{2}P^2  \right] -\frac{\sqrt{h}}{16\pi} R^{(3)}    \label{hamconstr}
\end{equation}
\begin{equation}
\mathcal{H} _i = -2 h_{an}D_c P^{cn}   \label{momconstr}
\end{equation}
The canonical momentum $ P^{ab}$ is defined as
\begin{equation}
\frac{1}{16\pi} \sqrt{h} ( K^{ab} -h^{ab}K )   \; .         \label{canmomentum}
\end{equation} 
Form (\ref{hamconstr}) and (\ref{momconstr}) we see that the bulk hamiltonian $H$ does not depend on the time derivatives of $N$ and $N^i$ and therefore the
conjugate momentum of those two quantities is zero. This means that we have 
\begin{equation}
\frac{\delta H}{\delta N} = \frac{\delta H}{\delta N^i} =0  \; .
\end{equation}
The constraints therefore are given by
\begin{equation}
\mathcal{H} = \mathcal{H} _i =0
\end{equation}
We want now to compute the constraint algebra using  the fundamental Poisson brackets
\begin{equation}
\left\{ P , P  \right\} = \left\{ h , h  \right\} =0       \label{fundamental1}
\end{equation}
and
\begin{equation}
\left\{ h_{nm} (x) , P^{ab}(x') \right\} = \delta ^a _n \delta ^b _m \delta (x-x') \label{fundamental2}
\end{equation}
we obtain the  constraint algebra in local form 
\begin{equation}
\left\{ \mathcal{H} _a (x) , \mathcal{H} _b (x') \right\} = \mathcal{H} _a (x') \partial _b \delta (x-x') + \mathcal{H} _b (x)\partial _a \delta(x-x')
\end{equation}
\begin{equation}
\left\{ \mathcal{H} _a (x) , \mathcal{H}(x')  \right\} = \mathcal{H}(x)\partial _a \delta (x-x')
\end{equation}
\begin{equation}
\left\{ \mathcal{H}(x) , \mathcal{H}(x')  \right\} = \left( \mathcal{H} ^a (x) + \mathcal{H} ^a(x')  \right)\partial _a \delta (x-x')
\end{equation}
Now in the symplectic formalism constraints $C^i$ are defined as functions on the phase space $\Gamma$ in the form $C^i : \Gamma \rightarrow R$.
The constraints as written in (\ref{momconstr}) and \ref{hamconstr}) are not functions as they depend on the coordinates of $\Sigma _t$.
In order to obtain functions on the phase space we must integrate them with some smearing functions.\\ 
Now in the symplectic formalism every function on the phase space defines a vector field by means of the symplectic gradient.
It is was shown in \cite{dirac} that those vector fields generate the gauge transformations. In the case of general relativity the gauge group 
is the diffeomorphism group and therefore the generators of diffeomorphisms associated to a vector
field $\xi$, using the decomposition orthogonal and tangent to $\Sigma _t$, are respectively
\begin{equation}
H _{\bot} [\xi] = \int \hat{\xi} ^{\bot} \mathcal{H} d^3x
\end{equation}
\begin{equation}
H _T [\xi ] = \int \hat{\xi} ^a \mathcal{H} _a d^3x  \; .
\end{equation}
Where $\hat{\xi} ^{\bot}$ and $\hat{\xi} ^a $ are the components of $\xi$ normal and tangent to the hypersurface $\Sigma _t$.\\
We want now to compute the algebra of the integrated constraints using (\ref{fundamental1}) and (\ref{fundamental2}) obtaining
for the tangent part
\begin{equation}
\left\{ H_T [\xi ] , H _T[\eta ]  \right\} = H_T \left[ \hat{\xi}^T , \hat{\eta}^T \right] \; , 
\end{equation} 
where the bracket $\left[ \hat{\xi}^T , \hat{\eta}^T \right]$ is defined as
\begin{equation}
\left[\hat{\xi}^T , \hat{\eta} ^T     \right]^a = \hat{\xi} ^c \partial _c  \hat{\eta}^a - \hat{\eta} ^c \partial _c \hat{\xi} ^a
\end{equation}
For the mixed bracket we obtain
\begin{equation}
\left\{ H_T [\xi ] , H_{\bot} [\eta ]  \right\} = H_\bot [\hat{\xi} ^T \cdot \hat{\eta}^{\bot} ]
\end{equation}
with
\begin{equation}
\hat{\xi} ^T \cdot \hat{\eta}^{\bot} = \hat{\xi} ^a \partial _a \hat{\eta}^{\bot}   \; .
\end{equation}
For the orthogonal bracket we obtain
\begin{equation}
\left\{ H_{\bot} [\xi ] , H_{\bot} [\eta]   \right\} = H_T [ \hat{\xi} ^{\bot} * \hat{\eta} ^{\bot} ]
\end{equation}
with
\begin{equation}
\left( \hat{\xi}^{\bot} * \hat{\eta}^{\bot}  \right) _a = \hat{\xi}^{\bot} \partial _a \hat{\eta}^{\bot} - \hat{\eta}^{\bot}\partial_a \hat{\xi}^{\bot} 
\end{equation}
In computing this brackets we have ignored the boundary terms. Depending on the form the smearing vector $\xi$ as usual boundary terms must be added 
to the bulk generators.\\
putting now all the Poisson brackets together we can now write
\begin{equation}
\left\{ H[\xi ] , H[\eta ]   \right\} = H[\xi , \eta ] _{SD} \; ,  
\end{equation}
where the surface deformation algebra $[ , ]_{SD}$ is defined as
\[
\left[\hat{\xi} , \hat{\eta}   \right] _{SD} ^{\bot} = \hat{\xi} ^a \partial _a \hat{\eta} ^{\bot} - \hat{\eta} ^a \partial _a \hat{\xi} ^{\bot}
\]
\begin{equation}
\left[ \hat{\xi} , \hat{\eta} \right]_{SD} ^a = \hat{\xi} ^b \partial _b \hat{\eta} ^a -\hat{\eta} ^b \partial _b \hat{\xi} ^a 
+ h^{ab}\left( \hat{\xi} ^{\bot} \partial _b \hat{\eta} ^{\bot} - \hat{\eta} ^bot \partial _b \hat{\xi} ^{\bot}  \right)   \; .
\end{equation}
Therefore the Poisson algebra of diffeomorphism generators does not give as usual a representation of the Lie algebra.
The surface deformation algebra contains the metric $h_{ab} $ and therefore it has structure functions instead of structure constants.
Such an algebra is called open algebra.\\
On the other side the one can check that for the surface deformation algebra the Jacobi identity is not valid and therefore it is not a Lie
algebra.

\backmatter

\end{document}